\documentclass[preprint,aps,amssymb,showpacs,superscriptaddress,nofootinbib]{revtex4}
\usepackage{accents}
\usepackage{graphicx}
\usepackage{bbold}
\newcommand{\Case}[2]{{\textstyle \frac{#1}{#2}}}

\newcommand{\sgn}{\mathrm{sgn}(e)}

\begin{document}
\preprint{IMSc/2011/10/13}

\title{Revisiting canonical gravity with fermions}

\author{Ghanashyam Date}
\email{shyam@imsc.res.in}
\affiliation{The Institute of Mathematical Sciences\\
CIT Campus, Chennai-600 113, INDIA.}

\begin{abstract} 
Fermions constitute an important component of matter and their
quantization in presence of dynamical gravity is essential for any
theory of quantum gravity. We revisit the classical formulation adapted
for a background free quantization. The analysis is carried out with the
Hilbert-Palatini form for gravity together with the Nieh-Yan topological
term which keeps the nature of Barbero-Immirzi parameter independent of
inclusion of arbitrary matter with arbitrary couplings.  With dynamical
gravity, a priori, there are two distinct notions of `parity'$-$
orientation reversing diffeomorphisms and improper Lorentz rotations.
The invariance properties of the action and the canonical framework are
different with respect to these and gravitational origin of parity
violation seems ambiguous.

\end{abstract}


\maketitle

\section{Introduction}\label{Intro}

Incorporation of fermions in the background free quantum theory gravity
has been discussed in the literature \cite{History}. In the formulation
of general relativity in terms of real SU(2) connection, Thiemann
discussed loop quantization of standard model fields \cite{ThiemannQSD}.
The fermions were treated in the second order form i.e. fermions couple
to gravity through the spin connection (torsion free Lorentz
connection). Perez and Rovelli returned to fermions in presence of the
Holst term and found that its coefficient playing the role of the
inverse of the Barbero-Immirzi parameter, $\gamma$, becomes classically
observable \cite{PerezRovelli}. Mercuri \cite{Mercuri} discovered that
with a further addition of suitable non-minimal fermionic couplings,
$\gamma$ can be made classically unobservable. He also noted that these
added terms (Holst plus non-minimal) can be expressed as the Nieh-Yan
topological term once the connection equations of motion are used. The
strategy of adding non-minimal couplings to keep $\gamma$ classically
unobservable was followed for N = 1, 2 and 4 supergravities also
\cite{Kaul}. Canonical analysis and loop quantization of fermions with
non-minimal couplings was discussed by Bojowald and Das
\cite{BojowaldDas}. It was subsequently realised that $\gamma$ will
automatically be classically unobservable provided it is the (inverse)
of the coefficient of the Nieh-Yan term (a total divergence) in the
Lagrangian density. Thus, instead of the Holst terms alone, if the
Nieh-Yan term (Holst + torsion$^2$ piece) is used in conjunction with
the Hilbert-Palatini, then for {\em arbitrary} matter and their
couplings, the $\gamma$ will drop out of the classical equations of
motion. Furthermore, it is possible to systematically derive the real
SU(2) Hamiltonian formulation from such an action
\cite{DateKaulSengupta}.  Fermions were also included in the canonical
analysis. A necessary condition for a topological origin of $\gamma$ is
thus satisfied. The canonical analysis leading to real SU(2) formulation
has since been extended to supergravities \cite{KaulSenguptaSugra} as
well as further inclusion of the other two topological terms namely the
Pontryagin and the Euler classes \cite{PerezRezende, KaulSenguptaTop}. 

Fermions are also tied with possible parity violations \cite{Freidel,
BojowaldDas}.  There are two distinct notions of `parity': one related
to orientation of the space-time manifold and one related to the
improper Lorentz transformation. Depending upon the definitions of the
basic canonical variables (with or without the $\sgn$ factors in this
work (section \ref{Parities})), the canonical framework and the action
are (non-)invariant under {\em one} of the notions of parity. These
possibilities are not distinguished in the previous works.  The work in
\cite{DateKaulSengupta} is a little incomplete in the constraint
analysis although the final results are correct. The constraint
expressions are also not in a form which is suitable for loop
quantization. This work seeks to fill these gaps.

We re-derive the real SU(2) formulation including a Dirac fermion. The
analysis is done using restricted fields corresponding to `time gauge'.
Elimination of the second class constraints, leads to the usual
formulation. When a Dirac fermion is included, the solution of the
second class constraint leads to non-trivial Dirac brackets between the
SU(2) connection and the fermions. One can however make natural shifts
in the definition of the connection to recover the canonical brackets.
This also simplifies the constraints. Four Fermi interaction terms
however survive in the Hamiltonian and are signatures of first order
formulation. In the second order formulation where fermions couple to
the torsion free connection, there are no terms quartic in the fermions.

The straightforward derivation also introduces factors of $\sgn :=
$~sign(det($e^I_{\mu}$)) (which equals $N$sign(det($V^i_a$)) in the time
gauge parametrization) in appropriate places. For instance, we are
naturally lead to the definitions: $E^a_i :=  \sgn \sqrt{q}V^a_i, ~
A^i_a := \sgn~K^i_a + \eta~\Gamma^i_a$ (eqn. (\ref{ADef})).  Under an
{\em improper} orthogonal transformation ($O(3)$) acting on the index
$i$, the triad changes sign and so does the $\sgn$ factor leaving
$E^a_i$ invariant.  This is as it should be since the index $i$ on $E$
represents adjoint representation of SO(3) while on $V$ it represents
the defining representation. For SO(3), both are equivalent but {\em not
for O(3)}.  Under an {\em inversion}, $\Lambda^i_{~j} = -
\delta^i_{~j}$, quantities in the defining representation change sign
while those in the adjoint don't. The same reasoning applies to the
definition of the connection. Now the connection is also even under
inversion. The $\sgn$ factor also change the behaviour of $E^a_i$ and
$A^i_a$ under the action of {\em orientation reversing diffeomorphisms}.
These factors of $\sgn$ however occur only in the intermediate
derivations, the final form of constraints and basic variables are
independent of these factors.

The paper is organized as follows. Section \ref{CanonicalFormulation}
deals with the Hamiltonian formulation for purely gravitational sector
starting with the Hilbert-Palatini plus the Nieh-Yan action. Here steps
in the analysis are given and a gap in the constraint analysis in
\cite{DateKaulSengupta} is filled in. The constraints are presented in
the more standard form. The sign factors which appear in the action
itself, are followed through in all expressions.  Section \ref{Fermion}
deals with inclusion of a single Dirac fermion. Constraint analysis as
well as simplification of constraints is given here. The section ends
with the final form of the canonical formulation in the standard
notation. Section \ref{Parities} discusses the two distinct notions of
parities and the invariance properties of the action and the canonical
framework.  Section \ref{Summary} contains brief concluding remarks.

\section{The Hamiltonian formulation}\label{CanonicalFormulation}
The starting point is a choice of (tensor/spinor) fields and a
corresponding generally covariant, local action on 4 dimensional
space-time $M \simeq \mathbb{R} \times \Sigma_3$. The next step is to
carry out a 3 + 1 decomposition to identify the Lagrangian which is a
function of (tensor) fields on $\Sigma_3$ together with their velocities
with respect to the chosen time coordinate. The fields whose velocities
appear in the Lagrangian are {\em potentially} the configuration space
variables while those without velocities appearing in the Lagrangian are
{\em Lagrange multipliers} whose coefficients will be {\em primary
constraints}. This Lagrangian leads to the {\em kinematical phase
space}. Now a constraint analysis a la Dirac is performed. If there are
second class constraints, one may hope to simplify the analysis by
solving the second class constraints. However, now one must use the
Dirac brackets. These may not have the canonical form for the remaining
variables (i.e. may not be Darboux coordinates) and a new choice of
variables may be necessary.  This is particularly relevant for
Lagrangians which are {\em linear } in the velocities such as the
Hilbert-Palatini-Nieh-Yan and the Dirac Lagrangians which typically do
have primary, second class constraints. The classical Hamiltonian
formulation is completed when the action is expressed in the Hamiltonian
form together with first class constraints.  We also have fields
coordinatizing the kinematical phase space (after the second class
constraints are eliminated) with the configuration space coordinates
identified.
\subsection{Pure gravity}\label{VacuumGravity}
We begin with the Lagrangian 4-forms built from the basic fields the
co-tetrad $e^I_{\mu}dx^{\mu}$ and the Lorentz connection
$\omega^{IJ}_{\mu}dx^{\mu}$.
\begin{eqnarray}
{\cal L}_{\mathrm{HP}}(e, \omega) & = & \frac{1}{2\kappa}\left[
\mathrm{sgn}(e) \frac{1}{2}{\cal E}_{IJKL}R^{IJ}(\omega)\wedge e^K
\wedge e^L\right] \label{HP}\\
{\cal L}_{\mathrm{NY}}(e, \omega) & = & \left[T^I(e, \omega)\wedge
T_I(e, \omega) - R_{IJ}(\omega)\wedge e^I\wedge e^J\right] \label{NY}\\
{\cal L}_{\mathrm{grav}} & := & {\cal L}_{\mathrm{HP}} +
\frac{\eta}{2\kappa}{\cal L}_{\mathrm{NY}}  \label{LGrav}
\end{eqnarray}
Here $R := d\omega + \omega\wedge\omega$ and $T := de + \omega\wedge e$
are the usual curvature and torsion 2-forms\footnote{Our conventions are
such that (i) $\eta = \gamma^{-1}$, our action matches with
Ashtekar-Lewandowski (AL) action including signs and factors. Our
identification of $K^i_a$ is same is that of AL while our $\Gamma^i_a$
is minus that of AL. To match with the standard notation we first change
$A \to \eta A$ and then set $\eta = \gamma^{-1}$. This results in
standard definition of $F^i_{ab}$, our ${\cal G}^i \to \gamma^{-1}{\cal
G}^i_{std}$, the diffeomorphism and the Hamiltonian constraints remain
unchanged. The symplectic form becomes $\gamma^{-1}E^a_i \partial_t
A^i_a$. Therefore the Poisson brackets also become the standard ones.}
and $\kappa = 8\pi\mathrm{G}$.  The factor of sgn($e$) is present
because only then the Hilbert-Palatini Lagrangian matches with the
$\sqrt{|g|} R(g)$. This arises from noting that determinant of the
co-tetrad is given by, $e = $sgn$(e)\sqrt{|g|}$.

A 3 + 1 decomposition can be expressed using,
\begin{eqnarray}
d ~ := ~ d_{\perp} + d_{\parallel} & := & dt\partial_t + dx^a\partial_a
\nonumber \\
e^I ~ = ~ e^I_{\perp} + e^I_{\parallel} & , & 
\omega^{IJ} ~=~ \omega_{\perp}^{\ IJ} + \omega_{\parallel}^{\ IJ}
~~~~\mathrm{etc}
\end{eqnarray}
which leads to,
\begin{eqnarray}
{\cal L}_{\mathrm{grav}} & = &
\frac{1}{2\kappa}\left[\Sigma^{IJ}_{\parallel\parallel}\wedge
d_{\perp}\left(\mathrm{sgn}(e)\tilde{\omega}_{\parallel IJ} -
\eta\omega_{\parallel IJ}\right) + 2\eta T^I_{\parallel\parallel}\wedge
d_{\perp} e_{\parallel I} + 2\eta e^I_{\perp}\wedge
D_{\parallel}T_{I\parallel\parallel} \right. + \nonumber \\
& & \omega_{\perp
IJ}\wedge\left\{D_{\parallel}\left(\mathrm{sgn}(e)\tilde{\Sigma}^{IJ}_{\parallel\parallel}
-~\eta \Sigma^{IJ}_{\parallel\parallel}\right) - 2\eta
e^I_{\parallel}\wedge T^J_{\parallel\parallel}\right\} \nonumber \\
& & 
+~e^I_{\perp}\wedge\left\{e^J_{\parallel}\wedge\left(\mathrm{sgn}(e)\tilde{R}_{IJ\parallel\parallel}
-~R_{IJ\parallel\parallel}\right)\right\} + \left.
\mathrm{surface~terms}\right] ~~~~~\mathrm{where,}
\end{eqnarray}
$\Sigma^{IJ} := e^I\wedge e^J$, $D := d + \omega\wedge$ is the Lorentz
covariant derivative and $e$ is the determinant of the co-tetrad,
assumed to be non-zero. The $\tilde{}$ denotes the Lorentz dual defined
for any quantity antisymmetric in a pair of Lorentz indices, $X^{IJ}$,
as, 
\[ 
\tilde{X}^{IJ} ~:=~ \frac{1}{2}{\cal E}^{IJ}_{~~KL} X^{KL}. 
\]

A 3+1 decomposition is carried out as usual by choosing a foliation
defined by a time function $T:M \to \mathbb{R}$ and a vector field
$t^{\mu}\partial_{\mu}$, transversal to its leaves. The vector field is
normalised by $t\cdot\partial T = 1$ so that the parameters of its
integral curves, serve as the time coordinate. Given such a
decomposition, we choose a parametrization of the tetrad and the
co-tetrad as,
\begin{eqnarray}
e^I_{t} & = & Nn^I + N^aV_a^I ~~,~~ e^I_{a}
~=~V^I_{a} ~~~,~~~ n^In_I = -1 ~~,~~ n^IV^J_{a}
\eta_{IJ} = 0 ~~; \label{CoTetrad} \\
e_I^{t} & = & - N^{-1} n_I ~~,~~ e^{a}_I ~=~ N^{-1} n_I
N^{a} + V_I^{a} ~~~,~~~ n^I V_I^a = 0~~, \nonumber \\
\mathrm{with} & & V^a_I V^J_a ~=~ \delta_I^J + n_I n^J ~~,~~ V^a_I V^I_b
= \delta^a_b~~. \label{Tetrad}
\end{eqnarray}

Expressing the wedge products in terms of the components, using
$dx^t\wedge dx^a\wedge dx^b\wedge dx^c := {\cal E}^{tabc}dx^t\wedge
dx^1\wedge dx^2\wedge dx^3$ and separating the $IJ$ sums into $(0i),
(jk)$ sums, we write the gravitational Lagrangian density as, 
\begin{eqnarray}
{\cal L}_{\mathrm{grav}} & = & \frac{{\cal E}^{tabc}}{2\kappa} \left[
2V^0_bV^i_c \partial_t\left(\mathrm{sgn}(e) \tilde{\omega}_{a0i} -
\eta\omega_{a0i}\right) + V^i_bV^j_c\partial_t\left(\mathrm{sgn}(e)
\tilde{\omega}_{aij}
- \eta\omega_{aij}\right) \right. \nonumber \\
& & + \omega_{t0i}\left\{D_a\left(\mathrm{sgn}(e)
\tilde{\Sigma}^{0i}_{bc} - \eta\Sigma^{0i}_{bc}\right) -
D_a\left(\mathrm{sgn}(e) \tilde{\Sigma}^{i0}_{bc} -
\eta\Sigma^{i0}_{bc}\right) \right\} \nonumber \\
& & + \omega_{tij}\left\{D_a\left(\mathrm{sgn}(e)
\tilde{\Sigma}^{ij}_{bc} - \eta\Sigma^{ij}_{bc}\right) \right\}
\nonumber \\
& & + N \left\{n^0\left(V^i_a\left(\mathrm{sgn}(e) \tilde{R}_{0ibc} -
\eta R_{0ibc}\right) \right) \right. + \nonumber \\
& & \left.~~~~~~~~ n^i\left( - V^0_a\left(\mathrm{sgn}(e)
\tilde{R}_{0ibc} - \eta R_{0ibc}\right) + V^j_a\left(\mathrm{sgn}(e)
\tilde{R}_{ijbc} - R_{0ibc}\right) \right)\right\} \nonumber \\
& & + N^{a'}\left\{V^0_{a^{'}}\left(V^i_a\left(\mathrm{sgn}(e)
\tilde{R}_{0ibc} - \eta R_{0ibc}\right)\right) + \right.\nonumber \\ 
& & \left. \left.~~~~~~~~~ V^i_{a^{'}}\left(-V^0_a\left(\mathrm{sgn}(e)
\tilde{R}_{0ibc}
- \eta R_{0ibc}\right) + V^j_a\left(\mathrm{sgn}(e) \tilde{R}_{ijbc} -
  \eta R_{ijbc}\right) \right)\right\}\right] \nonumber \\
& & + \frac{1}{\kappa} \left[{T}^{aI}\partial_t{V}_{aI} ~+
\omega_{t0i}\left\{- {V}^0_a {T}^{ai} + {V}^i_a {T}^{a0} \right\} +
~\omega_{tij}\left\{ - {V}^i_a {T}^{aj} \right\} \right. \nonumber \\
& & \left. + N\left\{n^0\left(D_a{T}^{a0}\right)  + n^i\left(D_a
{T}^a_{i}\right) \right\}
+ N^{a'} \left\{V^0_{a'}\left(D_a {T}^a_{0}\right)  + V^i_{a'}\left(D_a
{T}^a_{i} \right)\right\} \right]
\end{eqnarray}

The terms in the last two lines, come from the torsion piece of the
Nieh-Yan term.  We have also defined,
\begin{equation}
T^{aI} ~:=~ \frac{\eta}{2}{\cal E}^{abc}T^I_{bc} ~=~ \frac{\eta}{2}{\cal
E}^{tabc}\left(\partial_bV^I_c  - \partial_cV^I_b + \omega_{b~J}^{\
I}V^J_c
- \omega_{c~J}^{\ I}V^J_b\right)
\end{equation}

We will now restrict to configurations such that $n_i = 0, n_0 = -1$.
This also implies that $V_{a}^0 = 0 = V_0^{a}$ and that $V^a_i$ are
invertible with $V_a^i$ as the inverse\footnote{This would correspond to
the choice of the so-called {\em time gauge} if we started without
restricting the configurations a priori.}. We also define the 3-metric
$q_{ab} := V_a^i V_b^j \delta_{ij}$ (which is positive definite in
classical theory) and denote $q := $det($q_{ab}$). Many terms in the
above equation drop out. In particular, there isn't any time derivative
of $V_{a0}$ in the last square bracket. We also promote the torsion
components $T^{ai}$ to new independent variables $\hat{T}^{ai}$ and
$V_a^i$ to $\hat{V}_a^i$, for the terms in the last square bracket. The
number of variables is restored back by introducing {\em two primary
constraints:} $\upsilon_a^i := \hat{V}^i_a - V^i_a \approx 0$ and
$\tau^a_i := \hat{T}^a_i - T^a_i \approx 0$ with $\xi^a_i, \phi^i_a$ as
the corresponding Lagrange multipliers. Note that $T^{a0}$ is {\em not}
promoted to a new variable. This simplifies the Lagrangian to,
\begin{eqnarray}
{\cal L}_{\mathrm{grav}} & = & \frac{{\cal E}^{tabc}}{2\kappa} \left[
V^i_bV^j_c\partial_t\left(\mathrm{sgn}(e)\tilde{\omega}_{aij}
- \eta\omega_{aij}\right) \right. \nonumber \\
& & +
\omega_{t0i}\left\{D_a\left(\mathrm{sgn}(e)\tilde{\Sigma}^{0i}_{bc} -
\eta\Sigma^{0i}_{bc}\right) -
D_a\left(\mathrm{sgn}(e)\tilde{\Sigma}^{i0}_{bc} -
\eta\Sigma^{i0}_{bc}\right) \right\} \nonumber \\
& & +
\omega_{tij}\left\{D_a\left(\mathrm{sgn}(e)\tilde{\Sigma}^{ij}_{bc} -
\eta\Sigma^{ij}_{bc}\right) \right\} \nonumber \\
& & + N \left\{V^i_a\left(\mathrm{sgn}(e)\tilde{R}_{0ibc} - \eta
R_{0ibc}\right) \right\} \nonumber \\
& & \left. + N^{a'}\left\{ V^i_{a^{'}}
V^j_a\left(\mathrm{sgn}(e)\tilde{R}_{ijbc} - \eta R_{ijbc}\right)
\right\}\right] \nonumber \\
& & + \frac{1}{\kappa} \left[\hat{T}^{ai}\partial_t\hat{V}_{ai}
-\xi^a_i(\hat{V}^i_a - V^i_a) - \phi^i_a(\hat{T}^a_i - T^a_i)\right.
\nonumber \\ 
& &  + \omega_{t0i}\left\{\hat{V}^i_a {T}^{a0} \right\} +
\omega_{tij}\left\{ - \hat{V}^i_a \hat{T}^{aj} \right\} \nonumber \\
& & \left. + N\left\{\left(D_a{T}^{a0}\right)  \right\}
+ N^{a'} \left\{\hat{V}^i_{a'}\left(D_a {T}^a_{i} \right)\right\}
\right]
\end{eqnarray}

At this stage it is convenient to introduce new notations for certain
combinations of the components of the Lorentz connection as well as
those of the (co)tetrad. These are,
\begin{eqnarray}
E^a_i & := & e \left(e^t_0 e^a_i - e^t_i e^a_0\right)
\hspace{1.55cm}\Rightarrow ~ E^a_i ~ = ~ \mathrm{sgn}(e)\sqrt{q} V^a_i
\\
K^i_a & := & \omega_a^{\ 0i} \hspace{3.6cm},~~~ \Gamma^i_a ~ := ~
\frac{1}{2}{\cal E}^i_{~jk} \omega_a^{\ jk} \hspace{1.0cm} \Rightarrow
\\
\omega_{aij} & = & {\cal E}_{ijk}\Gamma^k_a ~~,~~~\tilde{\omega}_{a0i} ~
= ~ \Gamma^i_a ~~~,~~~ \tilde{\omega}_{aij} ~ = ~ {\cal E}_{ijl} K_a^l
\end{eqnarray}
and we have used $e = N \mathrm{det}(V^i_a) = N
\mathrm{sgn}(e)\sqrt{q}$. Note that $N > 0$ implies that sgn($e$) =
sgn(det($V^i_a$)) and therefore we continue to use sgn($e$) every where. 

With these, the first velocity term becomes,
$2E^a_i\partial_t\left(\mathrm{sgn}(e) K^i_a - \eta\Gamma^i_a \right)$
suggesting the identification of a canonical pair, $(A^i_a, E^a_i)$ with 
\begin{equation} \label{ADef}
A^i_a ~ := ~ \mathrm{sgn}(e) K^i_a - \eta \Gamma^i_a 
\end{equation}
The velocity terms become,
\begin{equation} \label{VelTerm}
\left({\cal L}_{\mathrm{grav}}\right)_{\mathrm{velocity}} ~ = ~
\frac{1}{\kappa}\left\{ E^a_i \partial_t A^i_a +
\hat{T}^a_i\partial_t\hat{V}^i_a \right\} 
\end{equation}

At this stage we have 10 Lagrange multiplier fields, $N, N^a,
\omega_t^{\ IJ}$ and 18 canonical variables, $A_a^i, E^a_i$ while the
co-tetrad and the Lorentz connection constitute 13 + 24 = 37 fields.
Nine more fields are yet `unaccounted'. If we define 
\[
\accentset{\eta}{\omega}_a^{\ IJ} := \mathrm{sgn}(e)\omega_a^{\ IJ} +
\eta \tilde{\omega}_a^{\ IJ} ~ ,
\]
we can see that, $A^i_a = \accentset{\eta}{\omega}_a^{\ 0i}$. 

The nine remaining variables, $M_{kl}, \zeta^k$, are identified through
the definition\footnote{It is possible to bypass this decomposition and
work directly with $K^i_a$ and $A^i_a$ as independent canonical
variables. The conjugate momenta of $M_{ij}, \zeta^i$ will be subsumed
by the conjugate momenta of $K^i_a$ \cite{KaulSenguptaTop}. We will
however continue to use these variables.},
\begin{equation}
\accentset{\eta}{\omega}_a^{\ ij} := \frac{\sgn}{2}( E_a^i\zeta^j -
E_a^j\zeta^i ) + \frac{1}{2} {\cal E}^{ijk}M_{kl}E^l_a ~~~,~~~ M_{kl} =
M_{lk} ~~,~~ E_a^i E_i^b = \delta_a^b ~~,~~ E_a^i E^a_j = \delta^i_j \ .
\end{equation}
The 18 components of $\accentset{\eta}{\omega}_a^{\ IJ}$ are organized
as the 9 $A_a^i$, the 3 $\zeta^i$ and the 6 $M_{kl}$. The components $K,
\Gamma$ of the Lorentz connection $\omega$ can be expressed in terms of
the $A, \zeta \mathrm{and} M$ using the inverse formula,
\[
\omega_a^{\ IJ} ~ = ~ \frac{1}{1 + \eta^2}\left(
\sgn\accentset{\eta}{\omega}_a^{\ IJ} - \frac{\eta}{2} {\cal
E}^{IJ}_{~~KL} \ \accentset{\eta}{\omega}_a^{\ KL}\right) \ .
\]

The $K, \Gamma$ are given in terms of the independent canonical
variables $A, \zeta, M$ as,
\begin{equation} \label{KGammaDef}
K^i_a ~ = ~ \frac{1}{1 + \eta^2}\left\{ \sgn A^i_a +
\frac{\eta}{2}\left( M^i_{\ j}E^j_a + \sgn {\cal E}^i_{\ jk}E^j_a
\zeta^k\right) \right\} ~~~,~~~ \Gamma^i_a = \frac{1}{\eta}( \sgn K^i_a
- A^i_a) .
\end{equation}

For later convenience, we eliminate $\Gamma$ in favour of $K, A$. We
get,
\begin{eqnarray}
{\cal L}_{\mathrm{grav}} & = & \frac{1}{\kappa}\left[ E^a_i \partial_t
A^i_a + \hat{T}^a_i\partial_t\hat{V}^i_a \right. \nonumber \\ 
& & \left. ~~~ - \xi^a_iv^i_a - \phi^i_a \tau^a_i 
- \Lambda_i{\cal G}^{0i} - \lambda^k{\cal G}_k - N{\cal H} - N^a {\cal
  H}_a \right]  ~~, \label{GravL}\\
\kappa v^i_a & := & \hat{V}^i_a - V^i_a ~~, \\
\kappa \tau^a_i & := & \hat{T}^a_i - T^a_i ~~, \\
\kappa {\cal G}^{0i} & := & - \frac{1}{2}{\cal
E}^{tabc}\left[D_a\left(\sgn\tilde{\Sigma}^{0i}_{bc} - \eta
\Sigma^{0i}_{bc}\right) - D_a\left(\sgn\tilde{\Sigma}^{i0}_{bc} -
\eta\Sigma^{i0}_{bc}\right)\right] - \hat{V}^i_a{T}^{a0} ~~, \\
\kappa {\cal G}_k &:=& \frac{1}{2}{\cal E}_{kij}{\cal G}^{ij} ~ := ~ -
{\cal E}_{kij}\left[\frac{{\cal
E}^{tabc}}{2}D_a\left(\sgn\tilde{\Sigma}^{ij}_{bc}
- \eta\Sigma^{ij}_{bc}\right) - \hat{V}^i_a\hat{T}^{aj} \right] ~~, \\
\kappa {\cal H}_{a'} & := & - V^i_{a'}\left[\frac{{\cal E}^{tabc}}{2}
V_a^j\left(\sgn\tilde{R}_{ijbc} - \eta R_{ijbc}\right) +
D_a\hat{T}^a_i\right] ~~, \\
\kappa {\cal H} & := & - \left[ \frac{{\cal
E}^{tabc}}{2}V^i_a\left(\sgn\tilde{R}_{0ibc} - \eta R_{0ibc}\right) +
D_a{T}^{a0}\right]
\end{eqnarray}

In anticipation, we will refer to the ${\cal G}^{0i}$ as the {\em boost
constraints}, ${\cal G}_k$ as the {\em rotation constraints}, ${\cal
H}_{a'}$ as the {\em diffeomorphism constraints} and ${\cal H}$ as the
Hamiltonian constraint although at this stage the interpretation of the
transformations generated by these is ambiguous. The other two terms in
the second line are the two additional primary constraints which
identify the canonical coordinates $\hat{T}, \hat{V}$ with the torsion
and the co-tetrad components respectively. We have also used
$\omega_{t0i} := \Lambda_i, ~ \omega_{tij} := {\cal E}_{ijk}\lambda^k$.
Straightforward algebra then leads to\footnote{We set $\kappa = 1$ for
notational convenience. It will be restored back later.},
\begin{eqnarray}
{\cal G}^{0i} & = & \sgn\partial_a E^{ai} - \sgn\zeta^i -
\hat{V}^i_a{T}^{a0} \label{BoostConstraint}\\
{\cal G}_k & = & \eta \partial_a E^a_k + {\cal E}_{kij}A^i_a E^{aj} +
{\cal E}_{kij}\hat{V}^i_a\hat{T}^{aj} \label{RotationConstraint}\\
{\cal H}_{a'} & := & \eta F^i_{a'b} E^b_i + \left(\frac{1 +
\eta^2}{\eta}\right)V^i_{a'}K_{ib} K_{jc} V^j_a {\cal E}^{bca} \nonumber
\\ 
& & - V^i_{a'}\left\{ \partial_a\hat{T}^a_i - K_{ai}{T}^a_0 +
\eta^{-1}{\cal E}_{ijk}\hat{T}^{aj}( \sgn K_a^k - A_a^k ) \right\} ~~~~~
\mathrm{where,} \label{Diffeo}\\
F^i_{ab} & := & \frac{\partial_a A^i_b - \partial_b A^i_a}{\eta} +
\frac{{\cal E}^i_{\ jk}A^j_a A^k_b}{\eta^2} ~~, \label{FDefn}\\
{\cal H} & := &  - (\partial_a{T}^a_0 - K^i_a \hat{T}^a_i ) +
\frac{1}{2}\frac{E^b_j E^c_k}{\sqrt{q}}\left[ {\cal E}^{jk}_{~l}
F^l_{bc} + \left(\frac{1 + \eta^2}{\eta^2}\right)\left\{(K^j_b K^k_c
- K^j_c K^k_b) \right. \right. \nonumber \\
& & ~~~~~ \left. \left. - \sgn{\cal E}^{jk}_{~l}\left(\eta(\partial_b
K^l_c -~\partial_c K^l_b) + {\cal E}^l_{\ mn}(A^m_bK^n_c - A^m_c
K^n_b)\right)\right\}\right] \label{Hamiltonian}
\end{eqnarray}

As there are no velocities of $M_{ij}, \zeta^k$ variables, we have their
conjugate momenta $\pi^{ij} \approx 0 \approx \pi_k$ as {\em additional
primary} constraints. In Dirac's terminology, the {\em total} Hamiltonian
will be the minus terms in the second line of eqn (\ref{GravL}) plus
linear combinations of $\pi_i \approx \pi_{ij} \approx 0$. All these are
to be preserved during an evolution and this leads to {\em secondary
constraints} and determination of some Lagrange multipliers. 

Consider the preservation of the constraints $\pi_i, \pi_{ij},
\upsilon^i_a, \tau^a_i \approx 0$. This leads to the equations:
\begin{eqnarray} \{\pi_k, H_{tot}\} & \approx 0 \approx &
\int\left[\Lambda_i\frac{\delta{\cal G}^{0i}}{\delta\zeta^k} +
\cdots\right] \label{PiEqn} \\
\{\pi_{kl}, H_{tot}\} & \approx 0 \approx &
\int\left[\Lambda_i\frac{\delta{\cal G}^{0i}}{\delta M^{kl}} +
\lambda^i\frac{\delta{\cal G}_{i}}{\delta M^{kl}} + N^a\frac{\delta{\cal
H}_{a}}{\delta M^{kl}} + N\frac{\delta{\cal H}}{\delta M^{kl}} -
\phi^i_a\frac{\delta{T^a_i}}{\delta M^{kl}} + \xi_i^a \times 0 \right]
\label{PijEqn} \\
\{\upsilon^i_a, H_{tot}\} & \approx 0 \approx & \int\{\upsilon^i_a,
\tau_j^b\}\phi^j_b + \{\upsilon^i_a\ , \int \Lambda_j{\cal G}^{0j} +
\lambda^j{\cal G}_j + N^b{\cal H}_b + N {\cal H} \} \label{UpsilonEqn}
\\
\{\tau_i^a, H_{tot}\} & \approx 0 \approx & \int\{\tau_i^a,
\upsilon^j_b\}\xi_j^b + \cdots \label{TauEqn}
\end{eqnarray}

Now we note that $\Case{\delta{\cal G}^{0i}}{\delta \zeta^k} =
\sgn\delta^i_k$ and $\{\upsilon^i_a(x), \tau^b_j(y)\} = \delta^i_j\
\delta^b_a\ \delta^3(x, y)\ (1 + \eta^2)^{-1}$. This means that equation
(\ref{PiEqn}) can be solved for $\Lambda_i$, equation (\ref{UpsilonEqn})
can be solved for $\phi^i_a$ and equation (\ref{TauEqn}) can be solved
for $\xi^a_i$. (This also means that $(\pi_i, {\cal G}^{0j})$ and $
(\upsilon^i_a, \tau^b_j)$ constraints form second class pairs.)
Furthermore, from the appendix, we note that the boost and the rotation
constraints are independent of $M^{kl}$ and therefore the first two
terms of eqn (\ref{PijEqn}) are zero. This equation will turn out to
give a secondary constraint. These statements remain true also in the
presence of fermionic matter. Let us use the abbreviation $\bar{H} :=
\int N^a{\cal H}_a + N {\cal H}$. Then, using equation
(\ref{UpsilonEqn}), we obtain,
\begin{equation}
\frac{\phi^i_a(x)}{1 + \eta^2} ~=~ - \left\{\upsilon^i_a(x), \int
\Lambda_i{\cal G}^{0i} + \lambda^i{\cal G}_i\right\}
- \{\upsilon^i_a(x), \bar{H}\}
\end{equation}
Now, 
$$ \{\upsilon^i_a, X\} = \{\hat{V}^i_a - V^i_a, X\} = \frac{\delta
X}{\delta \hat{T}^a_i} + \frac{\sgn}{\sqrt{q}}\left(\frac{V^i_aV^k_c}{2}
- V^i_c V^k_a\right)\frac{\delta X}{\delta A^k_c} \hspace{1.0cm} \forall
  ~ X .  
$$
It is easy to see that for $X = \int \Lambda\cdot{\cal G} +
\lambda\cdot{\cal G}$, the Poisson bracket with $\upsilon^i_a$ is {\em
weakly zero}, with or without matter (fermionic). Using these, the
equation (\ref{PijEqn}) becomes,
\begin{eqnarray}
0 & \approx & \frac{\delta \bar{H}}{\delta M^{kl}} + (1 +
\eta^2)\left(\frac{\delta{T}^a_i}{\delta
M^{kl}}\right)\left\{\frac{\delta}{\delta\hat{T}^a_i} +
\frac{\sgn}{\sqrt{q}}\left(\frac{V^i_aV^j_b}{2} -
V^i_bV^j_a\right)\frac{\delta}{\delta A^j_b}\right\}\bar{H}
\hspace{0.3cm},\\
\frac{\delta T^a_i}{\delta M^{kl}} & = & \frac{\eta\ \sgn}{2(1 +
\eta^2)}\left(\delta_{ik}V^a_l + \delta_{il}V^a_k -
2\delta_{kl}V^a_i\right) \hspace{2.0cm}\mathrm{and}\\
\frac{\delta\bar{H}}{\delta\hat{T}^a_i} & = & \partial_a(N^bV^i_b) +
{\cal E}^i_{~jk}V^j_bN^b \Gamma^k_a + N K^i_a \hspace{2.0cm}\mathrm{so
~that,}\\
0 & \approx & \left[\frac{\delta}{\delta M^{kl}} - \frac{\eta\
\sgn}{2}\left(E_{kb}\delta^j_l +
E_{lb}\delta^j_k\right)\frac{\delta}{\delta A^j_b}\right]\bar{H}
\label{MEqnOne}\\
& & \hspace{1.0cm} + \frac{\eta\ \sgn}{2}\left(\delta_{ik}V^a_l +
\delta_{il}V^a_k - 2\delta_{kl}V^a_i\right)\left\{\partial_a(N^bV^i_b) +
{\cal E}^i_{~jk}V^j_bN^b \Gamma^k_a + N K^i_a \right\} \nonumber
\end{eqnarray}

The second line of the last equation is independent of matter
contribution. These appear only in the first line of that equation.
Noting that the $M^{kl}$ dependence in $\bar{H}$ appears {\em only
through $K^i_a$}, we can trade derivative w.r.t. $M^{kl}$ to that with
respect to $K^i_a$. Furthermore, the $A^i_a$, dependence is both {\em
explicit} as well as {\em implicit through $K^i_a$}. This simplifies
equation (\ref{MEqnOne}) to,
\begin{eqnarray}
\left(E_{ak}\delta^i_l + E_{al}\delta^i_k\right) \frac{\hat{\delta}
\bar{H}}{\hat{\delta} A^i_a} & \approx & \left(\delta_{ik}V^a_l +
\delta_{il}V^a_k - 2\delta_{kl}V^a_i\right)\left\{\partial_a(N^bV^i_b) +
{\cal E}^i_{~jk}V^j_bN^b \Gamma^k_a + N K^i_a \right\}~~ \label{MEqnTwo}
\end{eqnarray}
On the left hand side, the $\hat{\delta}$ signifies that only the {\em
explicit} dependence on $A^i_a$ is to be picked up. The right hand side
is independent of matter contributions. The sgn factors cancel out.

A somewhat lengthy but straight forward computation yields, 
\begin{eqnarray}
0 & \approx & \frac{N}{\eta^2}\frac{S_{kl}}{\sqrt{q}} +
\frac{N^a}{\eta}\left(E_{ak}{\cal G}_l + E_{al}{\cal G}_k\right)
\hspace{2.0cm}\mathrm{where,} \\
\frac{S_{kl}}{\eta} & := & \frac{1}{1 + \eta^2}\left\{\sgn(M_{kl} -
\delta_{kl}M^i_{~i}) - \eta( A_{ak}E^a_l + A_{al}E^a_k -
2\delta_{kl}A^i_aE^a_i)\right\} \nonumber \\
& & \hspace{1.0cm} + ~ ~{\cal E}^{abc}\left(V_{ak}\partial_b V_{cl} +
V_{al}\partial_b V_{ck}\right) \label{SDefn}
\end{eqnarray}
All the dependence on the derivatives of the lapse and shift variables
disappears. The secondary constraint is just $S_{kl} \approx 0$: 
\begin{eqnarray}
%
0 & = & \frac{\eta}{1 + \eta^2}\left\{ \sgn M^{ij} -
\eta\left(A^i_aE^{aj} + A^j_a E^{ai}\right) - \delta^{ij} \left( \sgn
M^k_{\ k} - 2 \eta A^i_a E^a_i \right) \right\} \nonumber \\
& & \hspace{1.5cm} + ~ \eta {\cal E}^{abc}\left(V^i_a\partial_b V^j_c +
V^j_a\partial_b V^i_c\right)  ~\hspace{2.0cm}\Rightarrow \label{SijEqn}
\\
\sgn M^{ij} & = & - (1 + \eta^2){\cal E}^{abc}\left\{V^i_a \partial_b
V^j_c + V^j_a \partial_b V^i_c - \delta^{ij} V^k_a\partial_b
V_{ck}\right\} + \eta\left( A_a^i E^{aj} + A_a^j E^{ai} \right)
\label{MSoln}
\end{eqnarray}
When fermionic matter is added, this constraint gets modified and is
discussed in the next section.

There are no further tertiary constraints.

From eqn. (\ref{PiEqn}), we notice that $(\pi_k, {\cal G}^{0l})$ forms a
second class pair of constraints, one of which is a canonical variable.
Defining a Dirac bracket relative to this pair will allow us to set
these constraints strongly equal to zero. {\em Because} $\pi_k$ is a
canonical variable, the Dirac brackets of variables other than $\pi_k,
\zeta^l$, among themselves, coincide with the corresponding Poisson
brackets. This follows by noting (schematically),
\begin{eqnarray}
\Delta & := &  \{\pi, G(q^i, p_j, \pi, \zeta) \} \nonumber \\
\{f(q^i, p_j, \pi, \zeta), \ g(q^i, p_j, \pi, \zeta)\}_* & := & \{f, g\}
+ \{f, \pi\}\Delta^{-1}\{G, g\} - \{f, G\}\Delta^{-1}\{\pi, g\}
\nonumber \\
\Rightarrow ~~~ \{q^i, p_j\}_* & = & \{q^i, p_j\} ~~~,
\end{eqnarray}

Since there is no explicit $\pi_k$ dependence in any of the remaining
constraints, these expressions remain the same. We impose ${\cal G}^{0i}
= 0$ strongly and eliminate $\zeta^i = \partial_aE^{ai} - \hat{V}^i_a
T^{0a}$. From eqn. (\ref{TorsionEqns}) of the appendix, this sets $T^{ai} =
\Case{1}{2}S^{ij}V^a_j$.

Similarly, $(\pi_{kl}, S^{ij})$ form a second class pair hence we can
define Dirac brackets relative to these and impose these strongly. The
constraints $\pi_{kl}$ being canonical variables, the Dirac brackets
among the remaining canonical variables remain the same as their Poisson
brackets. Once again, setting $\pi_{ij} = 0$ changes no expressions
but $S^{ij} = 0$ gives $M^{ij}$ in terms of $A_a^i$ and $E^a_i$. Also
$S^{ij} = 0$ implies that $T^{ia} = 0$ and sets the $\tau^a_i =
\hat{T}^a_i \approx 0$.  

But now we can define Dirac brackets relative to the $(\upsilon_a^i,
\tau^b_j)$ second class pair. In this $\tau^a_i$ is now a canonical
variable and therefore the Dirac brackets among the remaining variables,
$A_a^i, E^b_j$, remain equal to their Poisson brackets.

When fermionic matter is included, the first two steps of elimination of
second class constraints will remain the same but with non-zero torsion
and this will lead to non-trivial Dirac brackets among the final set of
variables in the third step.

{\em Note:} Solving $S_{ij} = 0$ and ${\cal G}^{0i} = 0$ strongly,
determines $M_{ij}$ and $\zeta^k$ in terms of $A, E$. Substitution in
the expression for $\Gamma^i_a$ (eqn. \ref{KGammaDef}), leads to,
$$
\Gamma^i_a = -\frac{{\cal E}^{bcd}}{2}E_{aj}\big(V^i_b\partial_c V^j_d +
V^j_b\partial_c V^i_d  - \delta^{ij} V^k_b\partial_c V_{dk}\big) +
\frac{{\cal E}^i_{~jk}}{2} E^j_a \partial_b E^{bk}\ .
$$
Using the identities, $~{\cal E}^{ijk}E^a_k = {\cal E}^{abc}V^i_bV^j_c~$
and $~{\cal E}^{cab}(V^i_aV^j_b - V^i_b V^j_a) = {\cal E}^{cab}{\cal
E}^{ij}_{~k}{\cal E}^k_{~mn}V^m_aV^n_b~ $, one can see that this is
precisely the the usual expression for $\Gamma^i_a$ in terms of the
triad alone,
\begin{equation}\label{StdGamma}
\Gamma^i_a = \frac{{\cal E}^{ijk}}{2}V_k^b\left\{\partial_b V_{aj} -
\partial_a V_{bj} + V^c_j V^l_a \partial_b V_{cl}\right\}.
\end{equation}

Thus, in the pure gravity case, we get the Hamiltonian formulation in
terms of the $A_a^i, E^b_j$ variables with the original Poisson
brackets, we are left with only the rotation, the diffeomorphism and the
Hamiltonian constraints and the total Hamiltonian is made up of these
alone. It remains to simplify the expressions for these constraints.

\subsection{Simplification of Constraints}\label{SimplificationVacuum}

After the second class constraints are imposed, we get $T^{0a} = - \sgn\
\eta V^a_i{\cal G}^i \approx 0~$, $\hat{T}^{a}_i = T^a_i \approx 0$ and
$\hat{V}^i_a = V^i_a$. The boost constraint, eqn.
(\ref{BoostConstraint}), is strongly zero and the rotation constraint,
eqn (\ref{RotationConstraint}), takes the usual form: 
\begin{equation}
{\cal G}^i = \eta \partial_a E^{ai} + {\cal E}^i_{\ jk} A^j_a E^{ak}.
\label{RotationSimplified}
\end{equation}

Consider the diffeomorphism constraint. The second term in equation
(\ref{Diffeo}) simplifies as, 
\begin{eqnarray}
\frac{1 + \eta^2}{\eta}V^i_{a'}K_{ib}K_{jc}V^j_a{\cal E}^{bca} & = &
\left(\frac{1 + \eta^2}{\eta}\right) V^i_{a'}
K_{bi}K_{jc}\sgn\frac{E^b_kE^c_l}{\sqrt{q}}{\cal E}^{jkl} \nonumber \\
& = & \left(\frac{1 + \eta^2}{\eta}\right) V^i_{a'} \frac{\sgn\
K_{ib}{E^b_k}}{\sqrt{q}}\left(- {\cal E}^{jlk}K_{jb}E^b_l\right)
\nonumber \\
& = & - \left(\frac{1 + \eta^2}{\eta}\right) V^i_{a'} K_{ib}{V^b_k}\
\left[\frac{\sgn}{1 + \eta^2}\left(\eta\xi^k + {\cal
E}^{kjl}A_{bj}E^b_l\right)\right] \nonumber \\
& = & - \left(\frac{1 + \eta^2}{\eta}\right) V^i_{a'} K_{ib}V^b_k \
{\cal G}^k \sgn  \label{1stTerm} \\
\therefore {\cal H}_{a'} & = & \eta F^i_{a'b} E^b_i - \frac{1}{\eta}
V^i_{a'} K_{ib}V^b_k{\cal G}^k\ \sgn ~~~~\approx~~~~ \eta F^i_{a'b} E^b_i
\label{DiffeoSimplified}
\end{eqnarray}
In the last but one line we have used the equation (\ref{Rotn}). 

Next consider the Hamiltonian constraint (eqn.(\ref{Hamiltonian})),
after using the expressions for torsion. 
\begin{eqnarray}
{\cal H} & := &  \frac{1}{2}\frac{E^b_j E^c_k}{\sqrt{q}}\left[ {\cal
E}^{jk}_{~l}F^l_{bc}\right] -~\partial_a\left(\sgn\ \eta V^a_i{\cal
G}^i\right)  + \left(\frac{1 + \eta^2}{\eta^2}\right)
\left(\frac{1}{2}\frac{E^b_j E^c_k}{\sqrt{q}}\right)\left[ (K^j_b K^k_c
- K^j_c K^k_b) \right.  \nonumber \\
& & ~~~~~  \left. - \sgn\ {\cal E}^{jk}_{~l}\left\{\eta(\partial_b K^l_c
-~\partial_c K^l_b) + {\cal E}^l_{\ mn}(A^m_bK^n_c - A^m_c
K^n_b)\right\}\right] \label{HVac}
\end{eqnarray}
The $K$-dependent terms can be written as (after taking out the $(1 +
\eta^2)/(2\eta^2)$ factor) and using $\Case{E^b_jE^c_k{\cal
E}^{jk}_{~l}}{\sqrt{q}} = \sgn {\cal E}^{bca}V_{al}$,
\begin{eqnarray*}
& & -~2 \sgn\ \eta {\cal E}^{jk}_{~~l}\partial_b\left(\frac{E^b_j E^c_k
K^l_c}{\sqrt{q}}\right) + 2 \sgn\ \eta K_c^l {\cal E}^{abc}\partial_b
V_{al} \nonumber \\
& & -~2 \frac{\sgn}{\sqrt{q}}\ (A^i_aE^{am}) {\cal E}_{ijk}{\cal
E}^k_{~mn} (K^j_bE^{bn}) + \frac{\sgn}{\sqrt{q}}(K^i_aE^{am}) {\cal
E}_{ijk}{\cal E}^k_{~mn} (K^j_bE^{bn}) 
\end{eqnarray*}
The total derivative term simplifies as, 
\begin{eqnarray}
{\cal E}^{jkl}K_{cl}E^c_k & = & \frac{\sgn}{1 + \eta^2}\left[{\cal
E}^{jkl}A_{cl}E^c_k + \frac{\eta}{2}{\cal E}^{jkl}{\cal
E}_{lkm}\zeta^m\right] \nonumber \\
& = & - \frac{\sgn}{1 + \eta^2}\left[{\cal E}^{jkl}A_{ck}E^c_l +
\eta\zeta^j\right] ~~ = ~~ - \sgn\ {\cal G}^j \nonumber \\
\therefore 
- {\cal E}^{jk}_{~~l}\partial_b\left(\frac{E^b_j E^c_k
  K^l_c}{\sqrt{q}}\right) & = & \partial_b (\sgn\ V^b_i {\cal G}^i)
\end{eqnarray}
Using the (\ref{MSoln}) equation for $M_{ij}$ and the boost constraint
for eliminating $\zeta^i$, we get,
$$
\sgn  K^i_aE^{aj} - A^i_aE^{aj} ~ = ~ - \frac{\eta}{2}{\cal
E}^{abc}\left\{2 V^i_a\partial_bV^j_c -
\delta^{ij}V_a^k\partial_bV_{ck}\right\} \ .  
$$
Writing $
A^i_a E^{am} = (A^i_a E^{am} - \sgn K^i_a E^{am}) + \sgn K^i_a E^{am}
$ in the 3rd term,
the algebraic $K-$dependence simplifies to,
$$
- E^b_jE^c_k\left(K^j_bK^k_c - K^j_c K^k_b\right) \ .
$$
Combining all the terms, the Hamiltonian constraint takes the simplified
form,
\begin{eqnarray}
{\cal H} & := &  \frac{1}{2}\frac{E^b_j E^c_k}{\sqrt{q}}\left[ {\cal
E}^{jk}_{~l}F^l_{bc} - \frac{1 + \eta^2}{\eta^2} \left(K^j_b K^k_c -
K^j_c K^k_b\right) \right] + \frac{1}{\eta}\partial_a\left(\sgn
V^a_i{\cal G}^i\right) \label{HSimplified} .
\end{eqnarray}
The total derivative term differs from eqn. 2.24 of \cite{ALReview},
because that expression is derived from the Holst action while ours is
derived from the Hilbert-Palatini-Nieh-Yan action.

With this we recover the usual form of the constraints (in the time
gauge) starting from the Hilbert-Palatini action with Nieh-Yan terms
added. In the next sub-section we add a Dirac fermion {\em minimally}
coupled to gravity.

\section{Addition of a Dirac Fermion}\label{Fermion}

The Lagrangian we begin with is,
\begin{eqnarray}
{\cal L}_{\mathrm{Dirac}} & = & - \frac{i}{2} |e| \left[ \bar{\lambda}
e^{\mu}_I \gamma^I D_{\mu}(\omega, A, \ldots) \lambda -
{\overline{D_{\mu}(\omega, A, \ldots) \lambda}} e^{\mu}_I \gamma^I
\lambda \right] \\
D_{\mu}(\omega)\lambda & := & \partial_{\mu} \lambda +
\frac{1}{2}\omega_{\mu}^{\ IJ} \sigma_{IJ}\lambda + i e'A_{\mu}\lambda +
\ldots \lambda ~~~, \nonumber \\
\overline{D_{\mu}(\omega)\lambda} & := & \left\{\partial_{\mu}
\lambda^{\dagger} + \frac{1}{2}\omega_{\mu}^{\ IJ}
\lambda^{\dagger}\sigma^{\dagger}_{IJ} - i e' A_{\mu} \lambda^{\dagger}
+ \ldots \lambda^{\dagger} \right\}\gamma^0 
\end{eqnarray}
The $\ldots$ refer to possible couplings of the Dirac fermion to other
gauge fields eg the Maxwell field. These are suppressed in the
following. The conventions for the $\gamma^I-$ matrices (space-time
independent) are given in the appendix \ref{Notation}. The factor in
front of the Dirac Lagrangian is $-i$ because our metric signature is (-
+ + +). This is crucial for the Dirac brackets and subsequent passage to
quantization via `Quantum brackets = $i\hbar$ Dirac brackets' rule, with
the quantum brackets being realised on a Hilbert space.

For future convenience we introduce $\Psi := q^{1/4}\lambda,~
\Psi^{\dagger} := q^{1/4}\lambda^{\dagger}$. This absorbs away the
$\sqrt{q}$ factors in the Lagrangian as well as in the constraints. Note
that the terms involving the derivatives of $\sqrt{q}$ cancel out. The
$\lambda$ fermionic variables being of density weight zero, the $\Psi$
fermionic variables are of density weight 1/2. From now on we will use
the half density variables.

Substituting the 3 + 1 parametrization of the tetrad and using the
time-gauge, the Lagrangian can be written as,
\begin{eqnarray}
{\cal L}_{\mathrm{Dirac}} & = & \frac{i}{2} ( \Psi^{\dagger} \partial_t
\Psi - \partial_t (\Psi^{\dagger}) \Psi - \frac{1}{2} \omega_{tIJ} {\cal
G}^{IJ}_F - N^{a'} {\cal H}^F_{a'} - N {\cal H}_F ~~~ \mathrm{where,}
\nonumber \\
{\cal G}^{0i}_F & = & 0 \hspace{4.9cm},~~~~ {\cal G}^F_i  ~ = ~
- \frac{i}{2} {\cal E}_{ijk} \Psi^{\dagger}\sigma^{jk} \Psi \label{GiF}
  \\
{\cal H}^F_{a'} & = & - \frac{i}{2}\left( \bar{\Psi}\gamma^0 D_{a'} \Psi
- \overline{D_{a'} \Psi}\gamma^0 \Psi \right) ~~~,~~~
{\cal H}_F ~ = ~ \frac{i}{2} V^a_i \left( \bar{\Psi}\gamma^i D_a \Psi  -
\overline{D_a \Psi}\gamma^i \Psi \right)
\end{eqnarray}

As in the vacuum case, the fermionic contribution to the boost and the
rotation constraints is independent of $K$ and hence of $M$ and $\zeta$.
Hence, the $\zeta, {\cal G}^{0i}$ pair of second class constraint can be
eliminated in exactly the same manner as before leading to the same
determination of $\zeta$ and of course without affecting the Poisson
brackets among the remaining variables. Furthermore, as in the vacuum
case, the $\Lambda_i, \lambda^i$, terms drop out from the equation
(\ref{PijEqn}). The secondary constraint is thus determined from
equation (\ref{MEqnTwo}) with $\bar{H}$ now including the fermionic
contribution. We only need to pick out the {\em explicit} $A$
dependence.

This is easily done and leads to,
\begin{eqnarray}
\left(E_{ak}\delta^i_l + k \leftrightarrow
l\right)\frac{\hat{\delta}}{\hat{\delta} A^i_a}\bar{H}_F\vert_A & = &
\frac{i}{2\eta}\left[N^aE_{ak}{\cal E}_l^{~mn}\
\Psi^{\dagger}\sigma_{mn}\Psi + k \leftrightarrow l \right. \nonumber \\
& & \left. \hspace{0.7cm} - \frac{\sgn\ N}{2\sqrt{q}}\left\{{\cal
E}^{mn}_{~~l}\bar{\Psi}(\gamma_k\sigma_{mn} + \sigma_{mn}\gamma_k)\Psi +
k \leftrightarrow l \right\}\right] \\
& = & \frac{N^a}{\eta} \left( E_{ak}{\cal G}^F_l + E_{al}{\cal G}^F_k
\right) - \frac{\sgn\ N}{\eta\sqrt{q}} \left( \delta_{kl} \Psi^{\dagger}
\gamma_5 \Psi \right)
\end{eqnarray}
Adding these to the eqn. (\ref{MEqnTwo}) gives,
\begin{equation}
\frac{N}{\eta^2\sqrt{q}}\left[ S_{kl} - \sgn\ \eta\delta_{kl}
\Psi^{\dagger}\gamma_5\Psi \right] + \frac{N^a}{\eta}\left[E_{ak}{\cal
G}^{tot}_l + E_{al}{\cal G}^{tot}_k\right] ~ \approx ~ 0 
\end{equation}
Hence the secondary constraint becomes,
\begin{equation}
S_{ij} ~ \approx ~ \sgn\ \eta \delta_{ij} \Psi^{\dagger}\gamma_5\Psi ~~
=: S^F_{ij} \label{SijFDefn}
\end{equation}
and the solution for $M_{ij}$ becomes,
\begin{eqnarray}
\sgn\ M^{ij} & = & - (1 + \eta^2){\cal E}^{abc}\left\{V^i_a \partial_b
V^j_c + V^j_a \partial_b V^i_c - \delta^{ij} V^i_a\partial_b
V_{ci}\right\} + \eta\left( A_a^i E^{aj} + A_a^j E^{ai} \right)
\nonumber \\ 
& & - \frac{(1 + \eta^2)}{2} \ \sgn\ \delta^{ij} \Psi^{\dagger} \gamma_5
\Psi \ . \label{MwithPsiSoln}
\end{eqnarray}
After imposing the second class constraints strongly, we get, 
$$
\hat{T}^a_i ~ = ~ T^a_i ~ = ~ \frac{1}{2}S_{ij}V^{aj} ~ = ~  \sgn\
\eta\frac{V^a_i}{2}\Psi^{\dagger}\gamma_5\Psi
$$
and this leads to {\em non-trivial Dirac brackets}. While we could
simplify the remaining constraints as before, it turns out to be better
to check the Dirac brackets relative to the ($\upsilon^i_a, \tau^a_i$)
second class pair which suggests a new definition of $A$ and simplifies
the constraints considerably.

For the fermions, the action being linear in velocities, we have primary
constraints, $\pi_{\lambda} \sim \lambda^{\dagger},
\pi_{\lambda^{\dagger}} \sim \lambda$, which are second class. Also
these variables fail to be Darboux coordinates - do not have vanishing
Poisson brackets\footnote{strictly {\em Generalised Poisson brackets}
\cite{HennauxTeitelbaum}, due to the Grassmann nature of the fermions.}
with the gravitational variables due to the $\sqrt{q}$ factor. The shift
to $\Psi, \Psi^{\dagger}$ variables makes the matter and gravitational
variables Poisson-commute. Defining Dirac brackets relative to these
primary, second class constraints allows us to use $\Psi,
\Psi^{\dagger}$ as basic variables with Dirac brackets given
by\footnote{The minus sign in the basic Dirac bracket, is correlated
with the sign in the Lagrangian. Upon quantization, we will have
anti-commutator, $[\Psi, \Psi^{\dagger}]_+ \sim + \hbar$ which is
consistent with the Hilbert space inner product.},
\begin{equation}
\{ \Psi^{\alpha}(x), \ \Psi^{\dagger}_{\beta}(y) \}_+ ~ = ~
- i\delta^{\alpha}_{\beta} \delta^3(x, y) \label{FermionicBracket}
\end{equation}

Now we define Dirac brackets relative to the ($\upsilon, \tau$)
constraints: 
$$
\upsilon^i_a = \hat{V}^i_a - V^i_a, ~~~ \tau^a_i = \hat{T}^a_i
-~\frac{1}{2}S_{ij}V^{aj} = \hat{T}^a_i - \frac{\sgn\
\eta}{2}V^a_i\Psi^{\dagger}\gamma_5\Psi \ .
$$
$$
\{\upsilon^i_a, \upsilon^j_b\} = 0 = \{\tau^a_i, \tau^b_j\}~~,~~
\{\upsilon^I_a(x), \tau^b_j(y)\} = \delta^i_j\delta^b_a\delta^3(x, y)
$$
The Dirac bracket is defined as,
\begin{eqnarray}
\{f(x), g(y)\}^* & := & \{f(x), g(y)\} - \int dz \{f(x),
\upsilon^i_a(z)\}\{\tau^a_i(z), g(y)\} \nonumber \\ 
& & \hspace{2.3cm} + \int dz \{f(x), \tau_i^a(z)\}\{\upsilon_a^i(z),
g(y)\} \label{DiracBracketDefn} \ .
\end{eqnarray}

Due to the presence of the fermionic term in $\tau^a_i$, the Dirac
bracket between $A$ and the fermions is non-trivial while all other
basic Dirac brackets remain the same as the corresponding Poisson
brackets. Specifically,
\begin{eqnarray}
\{A^i_a(x), \Psi^{\alpha}(y) \}^* & = & - \sgn\ \frac{i\ \eta}{4} E^i_a
(\gamma_5\Psi)^{\alpha} \delta^3(x, y) ~~,~~ \nonumber \\
\{A^i_a(x), \Psi^{\dagger}_{\alpha}(y) \}^* & = &  + \sgn\ \frac{i\
\eta}{4} E^i_a (\Psi^{\dagger}\gamma_5)_{\alpha} \delta^3(x, y) \ .
\label{ModifiedBrackets}
\end{eqnarray} 
This suggests that if we use,
$$
\underline{A}^{i}_a ~ := ~ A^i_a - \sgn\ \frac{\eta}{4} E^i_a
\Psi^{\dagger}\gamma_5 \Psi \ ,
$$
all Dirac brackets become standard Poisson brackets. Thus, while
simplifying the constraints we will first effect the shift $A =
\underline{A} + \Case{\sgn\ \eta}{4}E\Psi^{\dagger}\gamma_5\Psi$. This results
in very natural expressions for the constraints.

The solution (\ref{MwithPsiSoln}) for $M^{ij}$ simplifies as,
\begin{equation}
M^{ij}(A, E, \Psi) ~ = ~ \underline{M}^{ij}(\underline{A}, E) -
\frac{1}{2}\delta^{ij}\Psi^{\dagger}\gamma_5\Psi ~~~~ \mathrm{where,~~}
\underline{M}(A, E) \mathrm{~is~defined~in~(\ref{MSoln})}\ ,
\end{equation}

while the $K, \Gamma$ simplify as,
\begin{equation}
K^i_a(A, E, \zeta) ~ = ~ \underline{K}^i_a(\underline{A}, E, \zeta)
~~~,~~~ \Gamma^i_a(A, E, \zeta) ~ = ~
\underline{\Gamma}^i_a(\underline{A}, E, \zeta) - \frac{\sgn}{4}E^i_a
\Psi^{\dagger}\gamma_5\Psi \label{ShiftedKGamma}
\end{equation}
The underlined functions are the functions obtained in the pure gravity
case (`torsion-free'). In particular,
$\underline{\Gamma}^i_a(\underline{A}, E, \zeta) =
\underline{\Gamma}(V)$ as given in equation (\ref{StdGamma}), when the
second class constraints are strongly solved. This is relevant for
regularization of the Hamiltonian constraint later on.

The shift of $A$ does not affect the expression for the boost constraint
(\ref{Boost}) since the extra term is killed by the ${\cal E}$. For the
same reason the rotation constraint (\ref{Rotation}) is also unaffected.

The diffeomorphism constraint simplifies as follows. Recall
eqn.(\ref{Diffeo}),
\begin{eqnarray}
{\cal H}_{a'} & = & \eta F^i_{a'b}E^b_i +
\frac{1}{\eta}V^i_{a'}A^k_a{\cal E}_{ijk}\hat{T}^{aj}
-V^i_{a'}\partial_a\hat{T}^a_i \nonumber \\
& & + \frac{1 + \eta^2}{\eta}V^i_{a'}K_{ib}K_{jc}V^j_a{\cal E}^{bca} -
V^i_{a'}K_{ai}T^{a0} - \frac{\sgn}{\eta}V^i_{a'}K^k_a{\cal
E}_{ijk}\hat{T}^{aj} \nonumber \\
& & - \frac{i}{2}\left\{\bar{\Psi}\gamma^0D_{a'}\Psi -
\overline{D_{a'}\Psi}\gamma^0\Psi\right\} \label{Diff}\\
\mathrm{where,} & & \nonumber \\
\hat{T}^a_i & = & \sgn\frac{\eta}{2}V^a_i\Psi^{\dagger}\gamma_5\Psi ~,~
T^{a0} =\ - \eta\ \sgn V^a_i{\cal G}_{vac}^i ~\approx~\sgn\eta V^a_i{\cal
G}^i_F ~,~{\cal G}^i_F ~=~ \Psi^{\dagger}\gamma_5\sigma_{0i}\Psi
\nonumber \\
D_{a'}\Psi & = & \left(\partial_{a'}\Psi - \frac{i}{\eta}A^i_{a'}~
\gamma_5\sigma_{0i}\Psi\right) + K^i_{a'}\left(1  +
\sgn\frac{i\gamma_5}{\eta}\right)\sigma_{0i}\Psi \nonumber \\
\overline{D_{a'}\Psi} & = & \left(\partial_{a'}\bar{\Psi} +
\frac{i}{\eta}A^i_{a'} ~ \bar{\Psi}\gamma_5\sigma_{0i}\right) -
K^i_{a'}~ \bar{\Psi}\left(1  +
\sgn\frac{i\gamma_5}{\eta}\right)\sigma_{0i}
\end{eqnarray}

In the above expression, we have separated the $K-$dependent terms and
also solved the second class constraints. 

The first step in simplification is to put $A^i_a = \underline{A}^i_a +
\sgn\Case{\eta}{4}E^i_a \Psi^{\dagger}\gamma_5\Psi$. The $K$ independent
pieces just have the $A$ dependence being replaced by ${\underline A}$
dependence with one additional term,
\begin{equation} \label{Extra}
\left[\mbox{$K-$indep. terms}\right](A) ~=~ \left[\mbox{$K-$indep.
terms}\right](\underline{A}) - \sgn\frac{\eta}{2}E^i_{a'}(\partial_a
E^a_i)(\Psi^{\dagger}\gamma_5\Psi)
\end{equation}
The $K-$dependent terms combine to give,
\begin{equation}
\left[\mbox{$K-$dependent terms}\right] ~ = ~ - \sgn
\frac{1}{2}E^i_{a'}{\cal G}_i^F(\Psi^{\dagger}\gamma_5\Psi)
\end{equation}
In getting this simplification, we have used the boost constraint, the
rotation constraint as well as $K = K(A, M, \zeta)$ expressions.

The middle term of the first line of equation (\ref{Diff}) and the last
term of equation (\ref{Extra}), combine to give
$-\sgn\Case{1}{2}E^i_{a'}\Psi^{\dagger}\gamma_5\Psi{\cal G}_i^{vac}$.
This in turn combines with the simplified $K-$dependent term to give a
term proportional to the {\em total} rotation constraint which is weakly
zero.  Thus, finally, we get the simplified diffeomorphism constraint
as,
\begin{eqnarray} 
{\cal H}_{a'}(\underline{A}, E, \Psi) & = & \eta
F^i_{a'b}(\underline{A})\  E^b_i - \frac{i}{2}\left\{
\bar{\Psi}\gamma^0{\cal D}_{a'}\Psi - \overline{{\cal
D}_{a'}\Psi}\gamma^0\Psi \right\} \hspace{2.0cm}
\mathrm{where,}\label{DiffFinal}\\
{\cal D}_{a'}\Psi & := & \left(\partial_{a'}\Psi - i\ \frac{{\underline
A}^i_{a'}}{\eta}~ \gamma_5\sigma_{0i}\Psi\right) ~~~,~~~
\overline{{\cal D}_{a'}\Psi} ~ := ~ \left(\partial_{a'}\bar{\Psi} - i\
\frac{\underline{A}^i_{a'}}{\eta} ~ \bar{\Psi}\gamma_5\sigma_{0i}\right) 
\end{eqnarray}
Notice that this is a simple additive form of the vacuum diffeomorphism and the
Dirac diffeomorphism constraint (coupled to the SU(2) gauge connection
$\underline{A}$). Four fermion terms generated by the second class
constraint have been neatly absorbed in the shifted $\underline{A}$.

Lastly, we simplify the Hamiltonian constraint. Recall
eqn.(\ref{Hamiltonian}),
\begin{eqnarray}
{\cal H} & = & \frac{1}{2}\frac{E^b_j E^c_k}{\sqrt{q}}{\cal
E}^{jk}_{~~l}F^l_{bc}(A) + \partial_a T^{a0} + \left(\frac{1 +
\eta^2}{\eta^2}\right)\left(\frac{E^b_k E^c_l}{2\sqrt{q}}\right)\Big\{
K^j_bK^k_c - K^j_cK^k_b   \nonumber \\
& & \hspace{0.0cm} \left. - \sgn{\cal
E}^{jk}_{~~l}\Big(\eta\big(\partial_bK^l_c - \partial_cK^l_b\big) +
{\cal E}^l_{~mn}\big(A^m_bK^n_c - A^m_cK^n_b\big)\Big)\right\} ~ + ~
K^i_a\hat{T}^a_i \\
& & + \frac{i}{2} V^a_i\left\{\bar{\Psi}\gamma^i D_a\Psi -
\overline{D_a\Psi}\gamma^i\Psi\right\} 
%
%
%
\end{eqnarray}
As before, second class constraints have been solved and the
$K-$dependent terms are separated. Substitute $A^i_a = \underline{A}^i_a
+ \sgn\Case{\eta}{4}E^i_a\Psi^{\dagger}\gamma_5\Psi$. Thanks to eqn.
(\ref{ShiftedKGamma}), we replace everywhere $K \to \underline{K}$.  The
$A-$dependent terms generate additional terms.
$$
\frac{1}{2}\frac{E^b_j E^c_k}{\sqrt{q}}{\cal E}^{jk}_{~~l}F^l_{bc}(A)
~~~~ \to \hspace{10.0cm}~\\
$$
$$ \hspace*{2.0cm}\frac{1}{2}\frac{E^b_j E^c_k}{\sqrt{q}}{\cal
E}^{jk}_{~~l}F^l_{bc}(\underline{A}) + \left\{\frac{{\cal
E}^{abc}}{4}V_a^l\partial_bE_{cl} +
\frac{\underline{A}^i_aV^a_i}{2\eta}\right\}\sgn\ \Psi^{\dagger}\gamma_5\Psi +
\frac{3}{16\sqrt{q}}\left(\Psi^{\dagger}\gamma_5\Psi\right)^2\ ,
$$

$$ 
-~\sgn\ \frac{1 + \eta^2}{\eta^2}\frac{E^b_jE^c_k}{2\sqrt{q}}{\cal
E}^{jk}_{~~l}{\cal E}^l_{~mn}\left(A^m_bK^n_c - A^m_cK^n_b\right) ~~~~
\to \hspace{4.5cm}~ 
$$
$$
\hspace*{1.0cm} - \sgn\ \frac{1 +
\eta^2}{\eta^2}\frac{E^b_jE^c_k}{2\sqrt{q}}{\cal E}^{jk}_{~~l}{\cal
E}^l_{~mn}\left(\underline{A}^m_bK^n_c - \underline{A}^m_cK^n_b\right) -
\frac{1 +
\eta^2}{2\eta\sqrt{q}}\underline{K}^i_aE^a_i\Psi^{\dagger}\gamma_5\Psi\
, 
$$

and,

$$
\frac{i}{2} V^a_i\left\{\bar{\Psi}\gamma^i D_a\Psi -
\overline{D_a\Psi}\gamma^i\Psi\right\}(A, K)  ~~~~ \to \hspace{6.5cm}~
$$
$$
\frac{i}{2} V^a_i\left\{\bar{\Psi}\gamma^i D_a\Psi -
\overline{D_a\Psi}\gamma^i\Psi\right\}(\underline{A}, \underline{K}) -
\frac{3}{8\sqrt{q}}\left(\Psi^{\dagger}\gamma_5\Psi\right)^2
\hspace{1.5cm}
$$

The $K-$dependent terns in the fermionic Hamiltonian and the torsion
pieces combine to give,
$$
- \sgn\ {\cal E}^k_{~ij}K^i_aE^{aj}\frac{{\cal G}^k_F}{\sqrt{q}} +
\sgn\ \partial_a\left(\eta V^a_i{\cal G}^i_F\right) + \frac{1 +
\eta^2}{2\eta\sqrt{q}}K^i_aE^a_i\left(\Psi^{\dagger}\gamma_5\Psi\right)
$$

The terms generated by the shift in $A$, simplify to,
$$
\frac{K_a^iE^a_i}{2\eta\sqrt{q}}\left(\Psi^{\dagger}\gamma_5\Psi\right)
- 
\frac{(1 + \eta^2)K_a^iE^a_i}{2\eta\sqrt{q}}\left(\Psi^{\dagger}\gamma_5\Psi\right)
- \frac{3}{16}\frac{\left(\Psi^{\dagger}\gamma_5\Psi\right)^2}{\sqrt{q}}
$$

Using ${\cal G}^k_F = {\cal G}^k_{tot} - {\cal G}^k_{vac}$ and dropping
the term containing ${\cal G}^k_{tot}$\ , leads to 
\begin{eqnarray}
{\cal H} & := &  \frac{1}{2}\frac{E^b_j E^c_k}{\sqrt{q}}\left\{ {\cal
E}^{jk}_{~l}F^l_{bc}(\underline{A}) + \frac{1 + \eta^2}{\eta^2} \left(
\underline{K}^j_b \underline{K}^k_c - \underline{K}^j_c
\underline{K}^k_b\right) \right\} - \partial_a\left(\eta\ \sgn
V^a_i{\cal G}^i_{vac}\right)  \nonumber \\
& &  - \frac{1 + \eta^2}{\eta^2}\frac{E^b_jE^c_k}{2\sqrt{q}} {\cal
E}^{jk}_{~l}\Big\{ \eta\left(\partial_b \underline{K}^l_c -~\partial_c
\underline{K}^l_b\right) + {\cal E}^l_{\
mn}\left(\underline{A}^m_b\underline{K}^n_c
- \underline{A}^m_c \underline{K}^n_b\right)\Big\}\sgn \nonumber \\
& & + \frac{i}{2}V^a_i\left\{ \bar{\Psi}\gamma^i{\cal D}_{a}\Psi -
\overline{{\cal D}_{a}\Psi}\gamma^i\Psi \right\}(\underline{A}) \\
& & -
\frac{3}{16}\frac{\big(\Psi^{\dagger}\gamma_5\Psi\big)^2}{\sqrt{q}} +
\left(\frac{1}{4\sqrt{q}}{\cal E}^{abc}V^l_a\partial_bV_{cl} +
\frac{1}{2\sqrt{q}}\frac{\underline{A}^i_aE^a_i}{\eta}\right)\Big(\Psi^{\dagger}\gamma_5\Psi\Big)\sgn
+~\frac{{\cal G}^k_F {\cal G}_{kF}}{\sqrt{q}} \nonumber
\end{eqnarray}

The first two lines of the above equation are exactly the same as the
Hamiltonian for vacuum case and therefore are simplified in exactly the
same way as before. In particular, the derivatives of $\underline{K}$
terms after partial integration, generates another total derivative
term, namely, $(\eta^{-1} + \eta)\partial_a(\sgn V^a_i {\cal
G}^i_{vac})$, which combines with the last term in the first line
leaving us with $\eta^{-1}\partial_a(\sgn V^a_i{\cal G}^i_{vac})$. The
$k-$dependent terms just produce minus the second term in the first
line as before.

The third line is the contribution from the Dirac Hamiltonian coupled
to the $\underline{A}$ field while the terms in the fourth line are the
extra term including 4-fermions terms. 

Using the relation,
$$
\left(\frac{1}{4\sqrt{q}}{\cal E}^{abc}V^l_a\partial_bV_{cl} +
\frac{1}{2\sqrt{q}}\frac{\underline{A}^i_aE^a_i}{\eta}\right) =
\frac{\sgn}{2\eta\sqrt{q}}\underline{K}^i_a E^a_i
$$
The final form of the Hamiltonian constraint is,
\begin{eqnarray}
{\cal H} & := &  \frac{1}{2}\frac{E^b_j E^c_k}{\sqrt{q}}\left\{ {\cal
E}^{jk}_{~~l}F^l_{bc}(\underline{A}) - \frac{1 + \eta^2}{\eta^2} \left(
\underline{K}^j_b \underline{K}^k_c - \underline{K}^j_c
\underline{K}^k_b\right) \right\} + \frac{1}{\eta}\partial_a\left(\sgn\
V^a_i{\cal G}^i_{vac}\right)  \nonumber \\
& & + \frac{i}{2}V^a_i\left\{ \bar{\Psi}\gamma^i{\cal D}_{a}\Psi -
\overline{{\cal D}_{a}\Psi}\gamma^i\Psi \right\}(\underline{A}) + \left[
\left(\frac{1}{2\eta \sqrt{q}}\underline{K}^i_aE^a_i\right)
\Big(\Psi^{\dagger}\gamma_5\Psi\Big)\right] \nonumber \\
& & - \left[
\frac{3}{16}\frac{\big(\Psi^{\dagger}\gamma_5\Psi\big)^2}{\sqrt{q}}
-~\frac{{\cal G}^k_F {\cal G}_{kF}}{\sqrt{q}} \right] ~~~,~~~
\mathrm{where}~~~{\cal G}^i_{F} = \Psi^{\dagger}\gamma_5\sigma_{0i}\Psi
\label{HamFinal}
\end{eqnarray}

The Hamiltonian constraint thus consists of additive combination of the
vacuum and the Dirac Hamiltonian (coupled to $\underline{A}$). However,
unlike the diffeomorphism constraint (\ref{DiffFinal}), {\em there are
the additional terms in the square bracket}. These extra terms contain
contributions quartic in the fermions as well as quadratic in fermions.
Notice that there are no explicit factors of $\sgn$ in the final
expressions. The $\underline{K}^i_a$ appears which is the same as in the
vacuum case and therefore the properties needed in using the Thiemann
identities in the quantization of the gravitational Hamiltonian
constraint continue to hold. We have also checked that if the connection
equations of motion for the {\em spatial components} are solved and the
solution for these torsion components are substituted back in the
action, the above Hamiltonian is recovered. 

We close this section by reverting to the standard notation eg
\cite{ALReview}.

For this, we first substitute $\underline{A}^i_a \to \eta A^i_a$ (and
$\underline{K}^i_a \to K_a^i$). This removes the factors of $\eta^{-1}$
from the definition of $F^i_{ab}$ given in eqn. (\ref{FDefn}). It also
extracts a common factor of $\eta$ from the rotation constraint, ${\cal
G}^i$. The diffeomorphism and the Hamiltonian constraints remains the
same except for removing the factor of $\eta^{-1}$ from the total
derivative term in the Hamiltonian constraint. The symplectic term,
$E^a_i\partial_t A^i_a \to (\eta E^a_i)\partial_t A^i_a$. And finally we
put $\eta = \gamma^{-1}$ and restore $\kappa$ (Recall the overall
$\kappa^{-1}$ factor in the Lagrangian density of eqn (\ref{GravL})).

Here are the final expressions: 
\begin{equation} \label{FinalDefns}
P^a_i := (\kappa\gamma)^{-1} E^a_i = (\kappa\gamma)^{-1}\sgn
V^a_i\sqrt{q}  ~~~;~~~ A^i_a = \gamma\sgn K_a^i - \Gamma^i_a(V).
\end{equation}
\begin{equation} \label{FinalPB}
\{ A^i_a(x)\ , P^b_j(y) \} ~ = ~ \delta^b_a\delta^i_j \delta^3(x, y)
~~~,~~~ \{\Psi^{\alpha}(x)\ , \Psi^{\dagger}_{\beta}(y) \}_+ ~ = ~ - i
\delta^{\alpha}_{\beta}\delta^3(x,y) ~~.
\end{equation}
\begin{equation} \label{FinalGauss}
{\cal G}_i ~ = ~ \partial_a P^a_i + {\cal E}_{ij}^{~k} A^j_a P^a_k -
\frac{i}{2}{\cal E}_{ijk}\Psi^{\dagger}\sigma_{jk}\Psi ~~~~~,~~~
\sigma_{jk} := \frac{1}{4}[\gamma_j, \gamma_k]; \\
\end{equation}
\begin{equation} \label{FinalDiffeo}
{\cal H}_a ~ = ~ F^i_{ab}P^b_i - \frac{i}{2}\left(\bar{\Psi}\gamma^0
{\cal D}_a \Psi - \overline{{\cal D}_a\Psi} \gamma^0 \Psi\right) ~~~,~~~
{\cal D}_a \Psi := \left(\partial_a -i A^i_a \gamma_5\sigma_{0i}\right)
\Psi; \\
\end{equation}
\begin{eqnarray}
{\cal H} & := &  \kappa\gamma^2 \frac{1}{2}\frac{P^b_j
P^c_k}{\sqrt{q}}\left\{ {\cal E}^{jk}_{~~l}F^l_{bc}({A}) - (1 +
\gamma^2) \left( {K}^j_b {K}^k_c - {K}^j_c {K}^k_b\right) \right\} {\bf
+ \gamma\partial_a\left(\sgn\ V^a_i{\cal G}^i_{vac}\right) } \nonumber
\\
& & + \frac{i}{2}\frac{\kappa\gamma\, \sgn P^a_i}{\sqrt{q}}\left\{
\bar{\Psi}\gamma^i{\cal D}_{a}\Psi - \overline{{\cal
D}_{a}\Psi}\gamma^i\Psi \right\}({A}) + \left[
\left(\frac{\kappa\gamma^2}{2 \sqrt{q}}{K}^i_aP^a_i\right)
\Big(\Psi^{\dagger}\gamma_5\Psi\Big)\right] \nonumber \\
& & - \left[
\frac{3}{16}\kappa\frac{\big(\Psi^{\dagger}\gamma_5\Psi\big)^2}{\sqrt{q}}
-~\kappa\frac{(\Psi^{\dagger}\gamma_5\sigma^{0i}\Psi)(\Psi^{\dagger}\gamma_5\sigma^{0}_{\
i}\Psi)}{\sqrt{q}} \right] \label{FinalHam}
\end{eqnarray}

{\em Note:} Dimensionally, $\kappa \sim L^2~,~ (A, K, \partial) \sim
L^{-1}~,~ E \sim L^0~,~ P \sim L^{-2}~,~ (\Psi, \bar{\Psi}) \sim
L^{-3/2}~,~ {\cal G} \sim L^{-3}~,~ ({\cal H}_a, {\cal H}) \sim L^{-4}$.

The inverse square root of $q$ and $K_a^i$ appearing above are
manipulated exactly as in the vacuum case. As remarked earlier, the $A$
and $K$ above correspond to the vacuum case for which the Thiemann
identities hold.  Explicitly, the identities we would use are:
\begin{eqnarray}
\sgn {\cal E}^{bca} V^i_a ~ = ~ {\cal E}^{ijk}\frac{E^b_j
E_k^c}{\sqrt{\mathrm{det}(E^a_i)}} \hspace{1.0cm}& , & \hspace{1.0cm} q
:= (\mathrm{det}(V^i_a))^2 = \mathrm{det}(E^a_i)
\end{eqnarray}
\begin{eqnarray}
\kappa\gamma\frac{\sgn}{2}V^i_a(x) ~ = ~ \left\{{A}^i_a(x), \int d^3y
\sqrt{q}\right\} & \Rightarrow &  \nonumber \\
{\cal E}^{ijk}\frac{E^b_j E_k^c}{\sqrt{\mathrm{det}(E^a_i)}} & = &
\frac{2}{\kappa\gamma} {\cal E}^{bca} \left\{{A}^i_a(x), \int d^3y
\sqrt{q}\right\}\ \ .
\end{eqnarray}
\begin{eqnarray}
H_E(1) ~ := ~ \frac{\kappa\gamma^2}{2}\int \frac{P^b_j
P^c_k}{\sqrt{q}}{\cal E}^{jk}_{~l}F^l_{bc} & , & \overline{{K}} ~ := ~
\int d^3y \ \sgn {K}^i_a P^a_i ~~~ \Rightarrow \\
\overline{{K}} ~ = ~ (\kappa\gamma^3)^{-1}\left\{ H_E(1)~, \int d^3y
\sqrt{q} \right\}\ & ~~~,~~~ &
\sgn {K}^i_a(x) ~ = ~ \{{A}^i_a(x)~,~ {\overline{K}}\}
\end{eqnarray}
These identities suffice to derive a quantization the Hamiltonian
constraint from that of the `Euclidean Hamiltonian constraint' (the
first term in the Hamiltonian constraint) and of the volume operator. 

\section{Action of Constraints and their algebra}
It is easy to see that the gauge constraint generates correct gauge
transformation of the basic fields. Specifically, with ${\cal
G}(\Lambda) := \int_{\Sigma_3}d^3x \Lambda^i {\cal G}_i$,
\begin{eqnarray}
\{ A^i_a(x), {\cal G}(\Lambda)\} & = & - {\cal D}_a \Lambda^i ~~ = ~~ -
\partial_a \Lambda^i - {\cal E}^i_{~jk}A^j_a\Lambda^k \\
\{ P_i^a(x), {\cal G}(\Lambda)\} & = & {\cal E}_{ij}^{~k}\Lambda^j P^a_k
\\
\{ \Psi^{\alpha}(x), {\cal G}(\Lambda)\} & = & -i \Lambda^i
(\gamma_5\sigma_{0i}\Psi)^{\alpha} \\
\{ \Psi^{\dagger}_{\alpha}(x), {\cal G}(\Lambda)\} & = & +i \Lambda^i
(\Psi^{\dagger}\gamma_5\sigma_{0i})_{\alpha} 
\end{eqnarray}

If we compute the infinitesimal action of the ${\cal H}_a$ constraint on
the basic variables, we see that it equals the Lie derivatives of the
basic variables only up to an SU(2) gauge transformation.  We are
however free to modify the constraints by adding suitable combinations
of themselves.  So we {\em define} the diffeomorphism constraint as:
\begin{eqnarray}\label{TrueDiffeo}
{\cal C}(\vec{N}) & := & \int_{\Sigma_3}d^3x N^a {\cal C}_a ~~~~
\mathrm{with,} \nonumber \\
{\cal C}_a & := & {\cal H}_a - A^i_a {\cal G}_i ~ = ~ P^b_i\partial_a
A^i_b - \partial_b(A^i_a P^b_i) +
\frac{i}{2}\left(\Psi^{\dagger}\partial_a \Psi -
\partial_a\Psi^{\dagger}\cdot\Psi\right) 
\end{eqnarray}
which leads to the infinitesimal transformations,
\begin{eqnarray}
\{ A^i_a(x), {\cal C}(\vec{N})\} & = & {\cal L}_{\vec{N}} A^i_a ~~ = ~~
\partial_a(N^bA^i_b) + N^b(\partial_b A^i_a - \partial_a A^i_b) \\
\{ P_i^a(x), {\cal C}(\vec{N})\} & = & {\cal L}_{\vec{N}} P_i^a ~~ = ~~
N^b\partial_b P^a_i - P_i^b \partial_b N^a + 1\cdot(\partial_b N^b)
P^a_i \\
\{ \Psi^{\alpha}(x), {\cal C}(\vec{N})\} & = & {\cal L}_{\vec{N}}
\Psi^{\alpha} ~~ = ~~ N^b\partial_b \Psi^{\alpha} +
\frac{1}{2}\cdot(\partial_b N^b)\Psi^{\alpha} \\
\{ \Psi^{\dagger}_{\alpha}(x), {\cal C}(\vec{N})\} & = & {\cal
L}_{\vec{N}} \Psi^{\dagger}_{\alpha} ~~ = ~~ N^b\partial_b
\Psi^{\dagger}_{\alpha} + \frac{1}{2}\cdot(\partial_b
N^b)\Psi^{\dagger}_{\alpha} 
\end{eqnarray}

This implies that $\{var, \int N^a C_a\} = {\cal L}_{N^a}(var)$ for all
variables. The Gauge constraint already generates the correct gauge
transformation of the basic variables. By inspection, it follows that
the gauge constraints (weakly) commute with the diffeomorphism and the
Hamiltonian constraint, the gauge constraint and the diffeomorphism constraints
form sub-algebras and the diffeomorphism constraint transforms the Hamiltonian
constraint by the Lie derivative. The non-trivial bracket is the bracket
of two Hamiltonian constraints.
%
%
%
\section{Parity and Internal Parity}\label{Parities}
Recall that we begin with the (co-)tetrad field $e^I_{\mu}$, the Lorentz
connection $\omega^{IJ}_{\mu}$ and the fermion fields $\lambda,
\bar{\lambda}$ (or $\Psi, \bar{\Psi}$) defined over a manifold $M \sim
\mathbb{R}\times\Sigma_3$ which is assumed to be orientable. With the
topology specified, $M$ can be taken to be {\em time-orientable} with
respect to all the metric tensors constructed by the parametrization
(\ref{CoTetrad},\ref{Tetrad}).  Obviously, $\Sigma_3$ is orientable as
well. 

There are two distinct sets of discrete transformations: orientation
reversing diffeomorphism of $M$ and a $O(1,3)$ transformation with
determinant = -1.  We will keep the time orientation fixed. Orientation
reversing diffeomorphism of $M$ will then be reversing the orientation
of $\Sigma_3$. We will refer to these as {\em parity} transformations.
The improper Lorentz transformations $\Lambda^I_J$, will also be taken
so that det$\Lambda = -1$ {\em and} $\Lambda^0_{~0} = 1$ and will be
referred to as an {\em Lorentz parity transformation}. 

After going to the canonical framework in the `time gauge', we have the
fields $A^i_a, P^a_i, \Psi, \bar{\Psi}$ defined on $\Sigma_3$. The
parity transformations are the orientation reversing diffeomorphism of
$\Sigma_3$ and the improper $O(3)$ transformation, {\em inversion} (say)
will be the Lorentz parity transformation. 

In the Lagrangian framework, the Hilbert-Palatini action (also the Euler
and the cosmological terms) are invariant under both sets of
transformations while the Nieh-Yan (as well as the Pontryagin and the
Holst) actions {\em change signs under parity} but are {\em invariant
under Lorentz parity.} Hence the combined action is {\em invariant under
Lorentz parity and non-invariant under parity.}

The variables of the canonical framework are defined in terms of those
of the Lagrangian framework. These definitions of the $SU(2)$ connection
in terms of $K$ and $\Gamma$ and the conjugate momentum in terms of the
triad {\em are} consistent with the $SO(3)$ gauge transformation
extended to include the Lorentz parity. Thus the triad which transforms
by the defining representation changes sign under Lorentz parity. The
`densitized triad' (or the conjugate momentum) transforms by the adjoint
representation and should be invariant under Lorentz parity. The $\sgn$
factor in their definitions precisely takes care of this. The same can
be seen in the definition of the connection. It is easy to see that the
symplectic structure and the constraints (vacuum) are {\em all invariant
under Lorentz parity}.

When fermions are included, these are {\em scalars} under diffeomorphism
and transform as $\Psi \to \gamma^0 \Psi$ under Lorentz parity. All the
constraints including fermions are invariant under Lorentz parity. This
is true in both the Lagrangian and the canonical frameworks.

With regards to parity the situation is different. The action is not
invariant under parity, due to the Nieh-Yan term. In the canonical
framework, the connection is not simply even/odd under parity since the
$K$ term changes sign while the $\Gamma$ does not. The `densitized
triad' also acquires an extra minus sign under parity (behaves as a
`pseudo-vector of weight 1'). The symplectic structure thus is {\em not}
invariant. The constraints also are not invariant under parity. This
is consistent with the non-invariance of the action. 

The action {\em is} invariant under parity combined with $\gamma \to -
\gamma$. Our definitions have the appropriate factors of $\gamma$ to
restore the simple (even) behaviour of the basic canonical variable
resulting also in the invariance of the Poisson brackets and
constraints.

In short, Lorentz parity is an invariance of the action as well as the
canonical framework and parity is not. However, parity combined with
$\gamma \to - \gamma$ is an invariance of both action and the canonical
framework. It is not our intention to suggest that the combined
operation be a physical symmetry (which depends on the quantum theory),
but it is useful in checking the algebra.

One could try to change the definitions of the basic variables by
dropping the $\sgn$ factors. This will result in expressions which can
be obtained from the above by putting $\sgn = 1$. This will restore the
`densitized triad' to its usual density weight 1 vector density status
and the connection to its 1-form status.  The canonical framework is
then {\em invariant} under parity (without changing sign of $\gamma$).
The action however is still {\em not} invariant. 

Under Lorentz parity, the connection does not have simple behaviour and
the conjugate momentum will be odd. This results in non-invariance of
the canonical framework.  The action however is still invariant under
Lorentz parity. If the sign of $\gamma$ is changed along with the
Lorentz parity transformation, then the basic variables are even, the
symplectic structure is invariant and so are the constraints {\em and
the action}. Thus, the definitions without the $\sgn$ factors,
interchanges the role of Lorentz parity and parity appropriately
combined with $\gamma \to - \gamma$.

With two different definitions (with and without the $\sgn$ factors), we
can ensure either invariance of the canonical framework under parity or
under Lorentz parity and this is independent of {\em minimally coupled}
fermions. The action however unambiguously remains invariant under
Lorentz parity and non-invariant under parity. Which of these is more
appropriate?

Observe that if we were to consider formulation in terms of the metric
tensor, then the notion of Lorentz parity is not even definable as there
is no internal Lorentz transformation. On the other hand, existence of
fermions (spinorial fields) requires an orientable manifold and using
the tetrad formulation making the Lorentz parity notion available. If
the orientation of the manifold is regarded as a fixed background
structure, then parity transformation is excluded by definition and
Lorentz parity alone is available. Which of these is relevant from an
observational point of view is not very clear and so also the issue of
`parity violation' via gravitational interactions.
\section{Concluding remarks}\label{Summary}
The next natural step is the loop quantization of the fermions with
interacting gravity. The kinematic Hilbert space of this system has been
already constructed \cite{ThiemannQSD}. The procedure given by Thiemann
can be followed in toto. The extra feature, not available in Thiemann's
discussion are the quartic terms in the fermionic sector. Their
regularization has been given in \cite{BojowaldDas} and we don't have
any thing new to add to this.

In summary, we have presented a canonical form of a Dirac fermion,
minimally coupled to the tetrad form of gravity including the Nieh-Yan
term. The Canonical analysis shows that the coefficient of the Nieh-Yan
term {\em is} the inverse of the usual Barbero-Immirzi parameter while
the topological nature of the Nieh-Yan term guarantees its
non-appearance in the classical equations of motion. One could consider
additional non-minimal couplings, not affecting the symplectic
structure, but these will not change the status of the Barbero-Immirzi
parameter. We saw the natural appearance of the $\sgn$ factors which
also serve to give definitions consistent with the two distinct notions
of parity and Lorentz parity. There is no Lorentz parity violation
either in the Lagrangian framework or the Hamiltonian framework for this
system. As noted earlier, among matter fields, fermions alone are
sensitive to the SU(2) action and contribute to the rotation (Gauss)
constraint. Its implications for the homogeneous, diagonal models has
already been explored by Bojowald and Das \cite{BojowaldDas}. For the
same reasons, fermions are likely mediators in the black hole
evaporation process. Due to the role of fermions in the chiral anomalies
in the usual quantization, they are also probes to see how loop
quantization does or does not accommodate chiral anomalies.  For
exploration and elaboration of these issues, it is necessary to have a
sufficiently precise control over the fermion-gravity system and
classical analysis is the first step in this direction.

\acknowledgments

The many discussions with Romesh Kaul are gratefully acknowledged. 

\newpage
\appendix
\section{}
\subsection{Notation and Conventions}\label{Notation}
The basic fields are the co-tetrad $e^I_{\mu}$, the Lorentz connection
$\omega^{IJ}_{\ \mu}$ and a Dirac spinor $\lambda^{\alpha}$. The
co-tetrad is taken to be non-degenerate and defines a metric $g_{\mu\nu}
:= e_{\mu}^I e_{\nu}^J \eta_{IJ}, ~ \eta_{IJ} = \mathrm{diag}( -1, 1, 1,
1)$. The general coordinate indices, $\mu, \nu \ldots $, are
raised/lowered with $g_{\mu\nu}, ~ g^{\mu\nu}$ while the Lorentz
indices, $I, J, \ldots $, are raised/lowered with $\eta_{IJ}, ~
\eta^{IJ}$.
\begin{eqnarray}
e  :=  \mathrm{det} (e^I_{\mu}) & := & \frac{1}{4!} {\cal
E}^{\mu\nu\alpha\beta}{\cal E}_{IJKL} e^I_{\mu} e^J_{\nu} e^K_{\alpha}
e^L_{\beta}  ~~~,~~~ {\cal E}^{tabc} = 1 = {\cal E}_{0123} ~~;\\
e{\cal E}^{IJKL} & = & - {\cal E}^{\mu\nu\alpha\beta}e^I_{\mu} e^J_{\nu}
e^K_{\alpha} e^L_{\beta} \\
2e\Sigma^{\mu\nu}_{IJ} & := & e(e^{\mu_I} e^{\nu}_J - e^{\nu_I}
e^{\mu}_J) ~=~ \frac{1}{2} {\cal E}^{\mu\nu\alpha\beta}{\cal E}_{IJKL}
e^K_{\alpha} e^L_{\beta} \\
\accentset{\sim}{X}^{IJ} & := & \frac{1}{2}{\cal E}^{IJ}_{~~KL} X^{KL}
~~~,~~~ \accentset{\sim}{\accentset{\sim}{X}} ~=~ - X \\
T^I_{\ \mu\nu} & := & D_{\mu}e^I_{\nu} - D_{\nu}e^I_{\mu} ~~~ , ~~~
D_{\mu}e^I_{\nu} ~ := ~ \partial_{\mu} e^I_{\mu} + \omega_{\mu~ J}^{\ I}
e^J_{\nu} ~~ , \\
R^{IJ}_{\ \mu\nu} & := & \partial_{\mu}\omega_{\nu}^{\ IJ} -
\partial_{\nu}\omega_{\mu}^{\ IJ} + \omega_{\mu~K}^{\ I} \omega_{\nu}^{\
KJ} - \omega_{\nu~K}^{\ I} \omega_{\mu}^{\ KJ} ~~, \nonumber \\
& & \nonumber\\
2\eta^{IJ}\mathbb{1} & = & \gamma^I\gamma^J + \gamma^J\gamma^I
~~,~~\sigma^{IJ} ~:=~\frac{1}{4}\left[ \gamma^I,
\gamma^J\right]~~,~~\gamma_5 ~:=~i\gamma^0\gamma^1\gamma^2\gamma^3 \\
\bar{\lambda} & := & \lambda^{\dagger}\gamma^0 \hspace{1.7cm},~~~
\gamma_I^{\dagger} ~=~ \gamma^0 \gamma_I \gamma^0 \hspace{1.1cm},~~~
{\cal E}_i^{~jk}\sigma_{jk} ~ = ~ 2 i \gamma_5 \sigma_{0i}~~,\\
D_{\mu}(\omega)\lambda & := & \partial_{\mu} \lambda +
\frac{1}{2}\omega_{\mu}^{\ IJ} \sigma_{IJ} \lambda ~~~,~~~ 
\overline{D_{\mu}(\omega)\lambda} ~ := ~ \left\{\partial_{\mu}
\lambda^{\dagger} + \frac{1}{2}\omega_{\mu}^{\ IJ}
\lambda^{\dagger}\sigma^{\dagger}_{IJ} + \right\}\gamma^0 ~. 
\end{eqnarray}
An explicit representation for the Dirac matrices is chosen to be,
\begin{equation}
\gamma^0 ~ := ~ \left( \begin{array}{cc} \mathbb{0} & - \mathbb{1} \\
\mathbb{1} & \mathbb{0} \end{array} \right) ~~~,~~~
\gamma^i ~ := ~ \left( \begin{array}{cc} \mathbb{0} & \sigma^i \\
\sigma^i & \mathbb{0} \end{array} \right) ~~~,~~~
\gamma_5 ~ := ~ \left( \begin{array}{cc} \mathbb{1} & \mathbb{0} \\
\mathbb{0} & - \mathbb{1} \end{array} \right) ~~~.
\end{equation}
\subsection{Torsion Components}\label{Torsion}
The torsion is defined geometrically as the covariant derivative of the
co-tetrad. These involve the components of the Lorentz connection which
have been obtained as functions of the canonical variables $A, E, M,
\zeta$. Similarly the boost and the rotation constraint are also
expressed as functions of $A, E, \zeta$. Eliminating $\zeta$ and some
combinations of ${\cal E}, A, E$, we can express the torsion in terms of
the constraints and the $\hat{V}, \hat{T}$ variables. Subsequently, we
will set $\hat{V} = V$ and $\hat{T} = T$ to simplify the expressions for
the torsion as well as the constraints.  The equations are,
\begin{eqnarray}
T^{a0} & := & \frac{\eta}{2} {\cal E}^{abc}T^0_{\ bc} ~ = ~ \eta {\cal
E}^{abc}K_{bi}V^i_c \\
T^{ai} & := & \frac{\eta}{2}{\cal E}^{abc}T^i_{\ bc} ~ = ~ \eta {\cal
E}^{abc} \left( \partial_b V^i_c + {\cal E}^i_{\ jk}\Gamma^k_b
V^j_c\right) \\
{\cal G}_{vac}^{0i} & = & \sgn(\partial_a E^{ai} - \zeta^i) -
\hat{V}^i_a{T}^{a0} \label{Boost}\\
{\cal G}^{vac}_k & = & \eta \partial_a E^a_k - {\cal E}_{kij}A^i_a
E^{aj} + {\cal E}_{kij} \hat{V}^i_a\hat{T}^{aj} \label{Rotation} 
\end{eqnarray}
Substitution of,
\begin{eqnarray}
{\cal E}^{abc} V^k_c & = & \sgn\sqrt{q} {\cal E}^{ijk} V^a_i V^b_j \\
K_{ai} E^a_j & = & \frac{1}{1 + \eta^2}\left[ \sgn A_{ai}E^a_j +
\frac{\eta}{2} M_{ij} + \frac{\eta}{2} \sgn {\cal E}_{ijk} \zeta^k
\right] \\
\Gamma_{ai} E^a_j & = & \frac{1}{1 + \eta^2}\left[ - \eta A_{ai}E^a_j +
\frac{\sgn}{2} M_{ij} + \frac{1}{2} {\cal E}_{ijk} \zeta^k  \right]
\end{eqnarray}
leads to, (using $E^a_i = \sgn\sqrt{q} V^a_i$ and $E_a^i = \sgn
V^i_a/\sqrt{q}$)
\begin{eqnarray}
T^{a0} & = & -\frac{\eta}{1 + \eta^2} V^a_i\ \sgn\left\{\eta\zeta^i +
{\cal E}^{ijk}A_{bj}E^b_k \right\} \nonumber \\
& = & - \eta V^a_i \left(\sgn{\cal G}_{vac}^i - \eta {\cal
G}_{vac}^{0i}\right) + \eta\ \sgn{\cal E}^i_{\
jk}V^{a}_i\hat{V}^j_b\hat{T}^{bk} \label{Ta0}\\
T^{ai}V^j_a + T^{aj}V^i_a & = & \hspace{0.0cm} \frac{\eta}{1 +
\eta^2}\left\{ \sgn\left(M^{ij} - ~\delta^{ij} M^k_{~k}\right) -
\eta\left(A^i_aE^{aj} + A^j_a E^{ai} - 2 \delta^{ij} A^k_a E^a_k\right)
\right\} \nonumber \\
&  &  \hspace{1.0cm} + ~ \eta {\cal E}^{abc}\left(V^i_a\partial_b V^j_c
+ V^j_a\partial_b V^i_c\right) \nonumber \\ 
& = & S^{ij} \hspace{6.0cm}[\mathrm{See\ eqn.(\ref{SDefn})}] \label{Sij}
\\
T^{ai}V^j_a - T^{aj}V^i_a & = & \eta {\cal E}^{ij}_{~ k}\left[
\partial_b E^{bk} + \frac{\eta}{1 + \eta^2} {\cal E}^k_{\ mn}A^m_bE^{bn}
- \frac{\zeta^k}{1 + \eta^2} \right] \nonumber \\
& = & \frac{\eta^2}{1 + \eta^2}\left(\hat{T}^{ai}\hat{V}^j_a -
\hat{T}^{aj}\hat{V}^i_a\right) + \nonumber \\
& & \hspace{1.0cm} \frac{\eta}{1 + \eta^2} \left[\sgn{\cal E}^{ij}_{~k}
V^k_a T^{a0} + {\cal E}^{ij}_{~k}\left(\eta {\cal G}_{vac}^k + \sgn{\cal
G}_{vac}^{0k}\right)\right] \nonumber \\
& = & \frac{\eta}{1 + \eta^2}{\cal E}^{ij}_{~ k}\left\{ \eta {\cal
E}^k_{\ mn} \hat{T}^m\hat{V}^n + \sgn V^k_a T^{a0} + \eta {\cal
G}_{vac}^k + \sgn {\cal G}_{vac}^{0k}\right\} \hspace{0.0cm}
\label{PreTai}\\
\therefore T^{ai}V_a^j & = & \frac{1}{2} S^{ij} +
\frac{\eta}{2}\sgn{\cal E}^{ij}_{~k}{\cal G}_{vac}^{0k} \label{Tai}
\end{eqnarray}
In the last line of the equation(\ref{PreTai}) we have used equation
(\ref{Ta0}). Now if we further put $\hat{T} = T, \hat{V} = V$, $T^{a0}$
simplifies to give,
\begin{equation}
T^{a0} ~ = ~ - \eta\ \sgn V^a_i {\cal G}_{vac}^i ~~~~~,~~~~~ T^{ai} ~ =
~ \frac{1}{2} S^{ij}V^a_j + \sgn\frac{\eta}{2}{\cal E}^{ij}_{~k} V^a_j
{\cal G}_{vac}^{0k} \label{TorsionEqns}
\end{equation}
Using these in the expressions of the boost (\ref{Boost}) and the Gauss
constraints (\ref{Rotation}), lead to,
\begin{eqnarray}
{\cal G}_{vac}^{0i} & = & \sgn \left[\partial_a E^{ai} -
\frac{\zeta^i}{1 + \eta^2} + \frac{\eta}{1 + \eta^2}{\cal E}^{i}_{\ jk}
A^j_a E^{ak}\right] \\
{\cal G}_{vac}^i & = & \frac{\eta}{1 + \eta^2} \zeta^i + \frac{1}{1 +
\eta^2}{\cal E}^{i}_{\ jk} A^j_a E^{ak} \label{Rotn}
\end{eqnarray}


\begin{thebibliography}{99}

\bibitem{History}
%
%
Jacobson, T, 1988, Fermions in canonical gravity, {\em Class. Quantum.
Grav.}, {\bf 5}, L143; \\
%
Morales-Tecotl, H A and Esposito, G, 1994, Selfdual action for fermionic
fields and gravitation, {\em  Nuovo Cim. B}, {\bf 109}, 973-982,
[gr-qc/9506073]; \\
%
Rovelli, C and Morales-Tecotl, H, 1995, Fermions in quantum gravity,
{\em Phys. Rev.  Lett.}, {\bf 72}, 3642-3645; \\
%
Rovelli, C and Morales-Tecotl, H, 1995, Loop space representation of
quantum fermions and gravity, {\em Nucl. Phys.  B}, {\bf 451}, 325-361;
\\
%
Baez, J, and Krasnov, K V, 1998, Quantization of
Diffeomorphism-Invariant Theories with Fermions {\em J. Math. Phys.},
{\bf 39}, 1251-1271, [arXiv:hep-th/9703112]\ . 
%
\bibitem{ThiemannQSD} 
%
Thiemann, T 1998, Kinematical Hilbert Spaces for Fermionic and Higgs
Quantum Field Theories, {\em Class.Quant.Grav.}, {\bf 15}, 1487-1512,
[arxiv:gr-qc/9705021]; \\
%
Thiemann, T 1998, QSD V : Quantum Gravity as the Natural Regulator of
Matter Quantum Field Theories, {\em Class.Quant.Grav.}, {\bf 15},
1281-1314, [arXiv:gr-qc/9705019]. 
%
\bibitem{PerezRovelli}
%
Perez, A and Rovelli, C, 2006, Physical effects of the Immirzi parameter
{\em Phys.Rev. D}, {\bf 73}, 044013, [arXiv:gr-qc/0505081]\ .
%
\bibitem{Mercuri}
%
Mercuri, S, 2006, Fermions in Ashtekar-Barbero Connections Formalism for
Arbitrary Values of the Immirzi Parameter, {\em Phys.Rev. D}, {\bf 73},
084016, [arXiv:gr-qc/0601013]\ ; \\
%
Mercuri, S, 2006, Nieh-Yan Invariant and Fermions in
Ashtekar-Barbero-Immirzi Formalism, [arXiv:gr-qc/0610026] 
%
\bibitem{Kaul}
%
Kaul, R K, 2008, Holst Actions for Supergravity Theories {\em Phys. Rev.
D}, {\bf 77}, 045030, [arXiv:0711.4674]\ .
%
\bibitem{BojowaldDas}
%
Bojowald M and Das, R, 2008, Canonical Gravity with Fermions, {\em Phys.
Rev. D}, {\bf 78}, 064009, [arXiv:0710.5722]; \\
%
Bojowald M and Das, R, 2008, Fermions in Loop Quantum Cosmology and the
Role of Parity {\em Class.Quant.Grav.}, {\bf 25}, 195006,
[arXiv:0806.2821] \ .
%
\bibitem{DateKaulSengupta}
%
Date, G, Kaul, R K and Sengupta, S, 2009, Topological Interpretation of
Barbero-Immirzi Parameter {\em Phys.Rev.D}, {\bf 79}, 044008,
[arXiv:0811.4496]\ .
%
\bibitem{KaulSenguptaSugra}
%
Sengupta, S and Kaul, R K, 2010, Canonical supergravity with
Barbero-Immirzi parameter {\em Phys.Rev.D}, {\bf 81}, 024024,
[arXiv:0909.4850]\ . 
%
\bibitem{PerezRezende}
%
Rezende, D J and Perez, A, 2009, 4d Lorentzian Holst action with
topological terms {\em Phys. Rev.}, {\bf D79}, 064026,
[arXiv:0902.3416]\ .
%
\bibitem{KaulSenguptaTop}
%
Sengupta, S and Kaul, R K, 2011, Topological parameters in gravity
[arXiv:1106.3027]\ .
%
\bibitem{Freidel} Freidel, L, Minic D and Takeuchi T, 2005, Quantum
Gravity, Torsion, Parity Violation and all that, Phys. Rev.  {\bf D72}
104002, [arXiv:hep-th/0507253]\ . 
%
\bibitem{ALReview}
%
Ashtekar,A and Lewandowski, J, 2004, Background Independent Quantum
Gravity: A Status Report {\em Class.Quant.Grav.}, {\bf 21}, R53,
[arXiv:gr-qc/0404018] 
%
\bibitem{HennauxTeitelbaum}
%
M Henneaux and C Teitelboim, 1992, {\em Quantization of Gauge Systems},
Princeton University Press, Princeton, New Jersey.

\end{thebibliography}
\end{document}